\documentclass[oldversion]{aa} 
\usepackage{graphicx}
\usepackage{txfonts}
\usepackage{enumerate}
\usepackage{natbib}
\bibpunct{(}{)}{;}{a}{}{,}
\begin{document}
   \title{The ultraluminous X-ray source NGC~1313 X-2}

   \subtitle{Its optical counterpart and environment}

   \author{F. Gris\'e
          \inst{1}
          \and
          M. W. Pakull
	  \inst{1}
	  \and
	  R. Soria
          \inst{2}
	  \and
	  C. Motch
	  \inst{1}
	  \and
	I. A. Smith 
	  \inst{3}
	  \and
	S. D. Ryder
	  \inst{4}
	  \and
	M. B\"ottcher
	  \inst{5}
	}


   \offprints{F. Gris\'e}

   \institute{Observatoire Astronomique de Strasbourg, 11 rue de l'Universit\'e, Strasbourg 67000, France
              \email{grise@astro.u-strasbg.fr}
	 \and
   Mullard Space Science Laboratory (UCL), Holmbury St Mary, Dorking, Surrey RH5 6NT
         \and
   Department of Physics and Astronomy, Rice University, 6100 South Main, MS‐108, Houston, TX 77251‐-1892
	 \and
   Anglo‐Australian Observatory, P.O. Box 296, Epping, NSW 1710, Australia
	 \and
   Astrophysical Institute, Department of Physics and Astronomy, Ohio University, Clippinger Hall 251B, Athens, OH 45701-‐2979
}
             

   \date{Received 11 February 2008 / Accepted 18 May 2008}

 \abstract{NGC~1313~X-2 is one of the brightest ultraluminous X-ray sources in
the sky, at both X-ray and optical wavelengths; therefore, quite a few
studies of available ESO VLT and HST data have appeared in the literature.
Here, we present our analysis of VLT/FORS1 and HST/ACS photometric data,
confirming the identification of the $B \sim 23$ mag blue optical counterpart. We show
that the system is part of a poor cluster with an age
of 20 Myr, leading to an upper mass limit of some 12 M$_\odot$ for the 
mass donor. We attribute the different results with respect to earlier studies to
the use of isochrones in the F435W and F555W HST/ACS photometric system that
appear to be incompatible with the corresponding Johnson B and V isochrones. 
The counterpart exhibits significant photometric variability of about 0.2 mag
amplitude, both between the two HST observations and during the one month of monitoring 
with the VLT. This includes variability within one night and suggests that
the light is dominated by the accretion disk in the system and not by the mass donor.}
 


   \maketitle
%

\section{Introduction}
\subsection{Context}

 Ultraluminous X-ray sources (ULXs) are extragalactic X-ray sources that are not at the nucleus of their galaxy, emitting
well above the Eddington limit of a $10\ \mathrm{M_{\sun}}$ black~hole ($L_\mathrm{X}\sim
10^{39}\ \mathrm{erg\ s^{-1}}$) if we assume that they emit isotropically. An important
question is whether they contain
intermediate mass black holes \citep{1999ApJ...519...89C,2000ApJ...535..632M}, whether they are 
beamed \citep{2001ApJ...552L.109K}, or if they are rather 
normal X-ray binaries with super-Eddington emission \citep{2002ApJ...568L..97B}.

The first clues about their compact nature came from X-ray studies showing that ULXs are variable on timescales of years down to minutes. Well-studied examples are Holmberg~IX~X-1 \citep{2001ApJ...556...47L}, ULXs in the Antennae \citep{2003ApJ...584L...5F} or M74~X-1 \citep{2005ApJ...630..228K}. Therefore, ULXs are accreting systems displaying spectra that were first described by a 
soft multicolor disk blackbody plus power-law continuum \citep{2003ApJ...585L..37M}. The soft component was interpreted as a cool accretion disk ($kT \sim 150$ eV). Using standard \citep{1973A&A....24..337S} multicolor disk models ($M \sim L^{1/2} T_{in}^{-2}$), this would imply black~hole (BH) masses $\sim 10^3$--$10^4\ \mathrm{M_{\sun}}$ assuming that the disk extends to the innermost stable circular orbit (as is thought to be the case in Galactic black-hole X-ray binaries in the disk-dominated high/soft state). However, it has been suggested \citep{2007ApSS.311..203R,2007Ap&SS.311..213S} that the assumptions might well be misleading: the observed energy distribution can be interpreted with alternative spectral models based on a cool outer disk and a Compton-scattering-dominated inner region. In this scenario, much lower masses ($\la 100\ \mathrm{M_{\sun}}$) are required for most ULXs, even in the absence of strong beaming.

Further clues to the ULX nature come from X-ray time-variability studies.
The detection of low-frequency quasi-periodic oscillations (LF-QPOs) in a
few ULXs exclude strong X-ray beaming \citep{2003ApJ...586L..61S,2006ApJ...641L.125D,2006MNRAS.365.1123M,2007ApJ...660..580S}. By
analogy with Galactic BHs and with some AGNs \citep{2006Natur.444..730M},
the characteristic frequencies of the LF-QPOs and of the breaks observed in
the power-density-spectra of a few ULXs suggest masses in the intermediate mass black hole (IMBH) range \citep{2007ApJ...660..580S,2007ApJ...663..445S}, if the varying
component is emitted at or near the innermost stable circular orbit.
However, as for the spectral interpretation, lower BH masses would be
implied if the oscillations come from a larger region, perhaps associated
with the thermal/non-thermal transition of the accretion flow. On longer
timescales (weeks to years), spectral state transitions have been observed
in a few luminous sources (e.g. \citealt{2001ApJ...556...47L,2004A&A...422..915S,2007MNRAS.379.1313S}), but it is not yet clear how they compare
with the ``canonical" state transitions in Galactic BHs (e.g. \citealt{2003ASPC..308..157R}).

Optical studies of ULXs have revealed point-like counterparts with blue
colors \citep{2002MNRAS.335L..67G,2002ApJ...580L..31L,2004MNRAS.351L..83K,2004ApJ...602..249L,2005MNRAS.356...12S,2005ApJ...620L..31K,2005ApJ...633L.101M,2006IAUS..230..302G,2007ApJ...661..165L,2008arXiv0803.3003R}, consistent with early-type donors (and more specifically,
with $10$--$20\  \mathrm{M_{\sun}}$ blue supergiants). However, the optical emission may have a
strong, perhaps dominant contribution from the accretion disk (which
should also have blue colors). This makes it more difficult to identify
the spectral type and mass of the donor star from the luminosity and
colors of the optical counterpart.

In many cases, optical counterparts are embedded in highly-supersonically expanding ionized nebul\ae\ (with possible contributions due to photoionization) and possibly related to the presence of jets \citep{2002astro.ph..2488P,2003RMxAC..15..197P,2004MNRAS.351L..83K,2007AstBu..62...36A,2007ApJ...668..124A,2008arXiv0803.4345P}. The study of nearby ULXs undergoing little interstellar extinction revealed that stellar counterparts are usually part of small young star clusters or OB associations \citep{2002ApJ...577..710Z,2002MNRAS.335L..67G,2006IAUS..230..302G,2007ApJ...661..165L,2007ApJ...668..124A}. But they are sometimes found slightly separated from these stellar associations \citep{2002ApJ...577..710Z}, as if they received a kick during a supernova event (which in this case could strengthen the stellar mass nature of the black hole). Radio counterparts are found in only few cases \citep{2005ApJ...623L.109M,2006MNRAS.368.1527S,2007ApJ...666...79L}. They are spatially resolved,
and have a spectrum consistent with optically-thin (steep) synchrotron
emission. Thus, they are likely to be jet-powered radio lobes, not core
emission from the collimated inner jet as observed in microquasars.

ULXs are mostly found in star-forming spirals or irregular galaxies 
\citep{2004ApJ...601L.143I,2004ApJS..154..519S}. 
Because of their extragalactic nature, these sources are optically faint. 
Telescopes with large collecting area such as the VLT and/or high spatial 
resolution such as HST are therefore required for the study of their optical counterparts 
and immediate environments.

\subsection{\object{NGC 1313 X-2} ULX}
Here, we study ULX X-2 in the barred spiral galaxy NGC
1313, which is located at a distance of about $4\ \mathrm{Mpc}$ 
($3.7\ \mathrm{Mpc}$ according to \citealt{1988ngc..book.....T}, $4.13\ \mathrm{Mpc}$ according 
to \citealt{2002AJ....124..213M}).
NGC~1313 has a mass of  M = 10$^{10.25}\ \mathrm{M_{\sun}}$ \citep{2004AJ....127.2031K}; 
this is roughly in the mass range where galaxies appear to be most efficient 
at producing ULXs \citep{2008AJsubmitted}.
X-2 is quite distant from the galactic nucleus, $\approx 6 \arcmin$ (corresponding to 
$7\ \mathrm{kpc}$) to the south. At this position, there are no
obvious signs of recent, extensive star formation. 
The galactic interstellar (IS) extinction is low ($E(B-V) \sim 0.1$, \citealt{1998ApJ...500..525S}): 
this makes X-2 one of the best targets for an optical spectroscopic 
and photometric study of a ULX and its host environment. 

\subsubsection{X-ray properties}

The source has been extensively studied in X-rays (see \citealt{2006ApJ...650L..75F} for a study
of 12 archival XMM-Newton observations, \citealt{2007PASJ...59S.257M} for a recent Suzaku study),
and has an average isotropic luminosity of $\approx 6\times 10^{39}\ \mathrm{erg s^{-1}}$ 
in the $0.3$--$10$ keV band. It was seen to vary in intensity by 
about $50\%$ in a few hours \citep{2007PASJ...59S.257M} and exhibits 
also significant variability on time scales of months and years. 
The X-ray spectrum of X-2 can be fitted
by at least 2 phenomenological models corresponding to 3 alternative physical mechanisms :
\begin{enumerate}[a)]
\item a hot thermal component ($kT_{\rm in} \approx 1.2$--$1.3$ keV) dominating above 2 keV, plus a 
soft down-scattered component. The hot thermal component was physically interpreted 
as a ``slim disk" around a stellar-mass black hole, at super-Eddington accretion rates \citep{2007PASJ...59S.257M}.
\item a cool thermal component ($kT_{\rm in} \approx 0.15$ keV) plus a power-law-like component dominating
above 2 keV. Physically, this can be interpreted in two ways :
\begin{enumerate}[i)]
\item as an optically-thick accretion disk extending to the innermost stable circular orbit, 
plus an optically-thin corona, around an intermediate-mass black hole
($M \sim 1000\ \mathrm{M_{\sun}}$), with accretion rates below 0.1 times Eddington \citep{2003ApJ...585L..37M}.
\item as an optically-thick disk, directly visible only far away from the innermost stable 
orbit, and replaced (or covered) at smaller radii by a hotter region ($kT \sim$ a few keV), 
optically-thick to electron scattering. This scenario may allow for less massive
black holes ($\la 100\ \mathrm{M_{\sun}}$) at accretion rates $\sim$ a few times Eddington
\citep{2007ApSS.311..203R,2007Ap&SS.311..213S}.
\end{enumerate}
\end{enumerate}
These three physical models predict different characteristic sizes, colors and luminosities for the accretion disk.
Optical studies can provide new constraints, if we can disentangle the optical contributions from the irradiated donor
star and the accretion disk.

\subsubsection{Optical properties}

At optical wavelengths, studies by \citet{2002astro.ph..2488P,2006IAUS..230..293P,2006ApJ...641..241R} revealed the presence of a 
huge ionized nebula (extension $18\arcsec \times 26\arcsec$, corresponding to $350 \times 500$~pc at a distance of 4.0 Mpc) at the position of X-2. The high expansion velocity of $\sim 100$ km/s \citep{2006ApJ...641..241R} underlines the suggestion that we are seeing emission from radiative shocks. This is supported by the presence of enhanced [OI] and [SII] forbidden lines (e.g. \citealt{2002astro.ph..2488P,2007AstBu..62...36A,2007ApJ...668..124A}). However, contributions due to photoionization by the X-ray source cannot presently be excluded.
The search for an optical counterpart of the ULX has motivated to carry out deep imaging of the field around it,
using the European Southern Observatory Very Large Telescope (ESO/VLT) 
and the Hubble Space Telescope Advanced Camera for Surveys (HST/ACS).
 
Multi-band photometry of the counterpart and of the surrounding stellar population
can constrain the nature of the donor star,
but results reported in the literature are quite discordant \citep[see][]{2005ApJ...633L.101M,
2007ApJ...658..999M,2006ApJ...641..241R,2007ApJ...661..165L}.
In particular, the age of the small cluster or association of relatively 
young stars around the ULX appears not to be well constrained. Determining this value is of great
interest, not at least because it tells us whether or not the ULX donor is a massive 
early-type star, and hence helps to constrain the physics and duration 
of the high mass-transfer phase.


In this paper, we analyze the full set of our photometric VLT observations
(some results were previously reported in \citealt{2006IAUS..230..293P}) 
together with archival HST/ACS data.
We will show that the ULX counterpart is part of a 20-Myr-old star cluster, 
which is not expected to contain uncollapsed stars more massive than 
$\approx 12\ \mathrm{M_{\sun}}$. 
Combining ground-based and HST data, we will present a detailed 
photometric light curve of the ULX counterpart, which shows large variability 
on timescales of hours and days. This suggests that the optical emission 
is dominated by an accretion disk with possible contributions by the X-ray-heated secondary.

\section{Observations}

\subsection{VLT Observations}

The first set of data comes from an observing programme (ID
072.D0614; PI: M. Pakull) carried out with the VLT/FORS1 instrument. 
Photometric monitoring was done over 9 nights and a spectroscopic 
study over 4 nights, between 2003 December 20 and 2004 January 15.
The field around X-2 was observed in broadband filters (standard 
$B, V, R$) and in some narrow-band filters (H$_\alpha$, [OIII]$\lambda$5007 and 
[OI]$\lambda$6300)(See Tables \ref{tab_vltdata} \& \ref{tab_vltdates} for details). 
We also monitored the source in the $B$ band with a total of 16 observations, 
each with an exposure time of $840\ \mathrm{s}$. Usually, we obtained one observation 
per night, except for 2003 December 24, when the field was observed 7 times. 
Each observation consisted of two dithered sub-exposures taken directly one after 
the other, and averaged to increase the signal-to-noise ratio.  
We also obtained two observations in $R$ and one in $V$ to study the color
of the source and of its stellar environment.
This paper mainly concentrates on photometric results; 
we leave the discussion of the spectroscopic results to a subsequent paper.

\begin{table*}
\caption{The VLT FORS1 observations for NGC~1313~X-2.}
\label{tab_vltdata}
\centering
\begin{tabular}{|l|c|c|c|c|}
\hline
Filter & Exposure Time (s) & Number of Exposures & Central Wavelength of Filter (\AA) & FWHM of Filter (\AA)\\
\hline
B      &     420       & 24		     &		4290	 &	880	\\
       &     840       & 4 		     &			 &		\\
V      &     600       & 2 		     &		5540	 &	1115	\\
R      &     500       & 4 		     &		6570	 &	1500	\\
H$_{\alpha}$ & 1500    & 2		     &		6563	 &	61	\\
$[$OI$]$ & 1300      & 2		     &		6295	 &	69	\\
$[$OIII$]$   & 1500      & 2		     &		5001	 &	55	\\
\hline

\end{tabular}
\end{table*}

\begin{table*}
\caption{Log of our VLT FORS1 broad-band observations.}
\label{tab_vltdates}
\centering
\begin{tabular}{|l|c|c|c|}
\hline
Filter & Observation Date & MJD at mid-exposure & Seeing ($\arcsec$)\\
\hline
B      & 2003 Dec. 21 & 52994.12864	& 0.60	\\
       & 2003 Dec. 22 & 52995.04826	& 0.56	\\
       & 2003 Dec. 23 & 52996.08222	& 0.66	\\
       & 2003 Dec. 24 & 52997.03231	& 0.58  \\
       &	      & 52997.08794	& 0.68  \\
       & 	      &	52997.09836	& 0.71	\\
       & 	      &	52997.12389	& 0.67	\\
       & 	      &	52997.20328	& 0.69	\\
       & 	      &	52997.21365	& 0.76	\\
       & 	      &	52997.24293	& 0.50	\\
       & 2003 Dec. 25 & 52998.17256	& 0.46	\\
       & 2003 Dec. 27 & 53000.22022	& 0.70	\\
       & 2003 Dec. 28 & 53001.17035	& 0.69	\\
       & 2003 Dec. 29 & 53002.21072	& 0.58	\\
       & 2003 Dec. 30 & 53003.04551	& 0.65	\\
       & 2004 Jan. 15 & 53019.04868	& 0.57	\\
V      & 2003 Dec. 25 & 52998.07801	& 0.55	\\
R      & 2003 Dec. 24 & 52997.11028 	& 0.75	\\
       & 	      & 52997.22568 	& 0.80	\\

\hline

\end{tabular}
\end{table*}

\subsection{HST Observations}

The second dataset was obtained from the HST/ACS archive 
(programme GO-9796; PI: J.~Miller). Observations were carried out 
on 2003 November 22, with the Wide Field Camera (WFC) 
in the F435W, F555W and F814W filters,
and with the High Resolution Camera (HRC) in the F330W filter. 
An additional observation was done on 2004 February 22 with 
the WFC in the F555W filter.
More detailed information about those exposures is summarized in Table~\ref{tab_hstdata}. We re-analyzed the data 
using the latest calibration files at the time of the analysis (STSDAS v3.5) and applied the standard correction 
for the spatial distortion (Multidrizzle 2.7.2, \citealt{2002hstc.conf..337K}).

\begin{table*}
\caption{The HST/ACS observations for NGC~1313~X-2.}
\label{tab_hstdata}
\centering
\begin{tabular}{|l|c|c|c|c|c|c|c|}
\hline
ID & Instruments & Filter & Date & MJD at mid-exposure  & Exposure Time (s) & FWHM of Filter (\AA)\\
\hline
j8ola2010 & HRC &  F330W  & 2003 Nov. 22	& 52965.44294	 &     2760    &	173.82 \\
j8ol02040 & WFC &  F435W  & 2003 Nov. 22	& 52965.37542	 &     2520    &	293.47 \\
j8ol02030 & WFC &  F555W  & 2003 Nov. 22	& 52965.31531    &     1160    &	360.02 \\
j8ol02010 & WFC &  F814W  & 2003 Nov. 22	& 52965.30656	 &     1160    &	654.64	 \\
j8ol06010 & WFC &  F555W  & 2004 Feb. 22	& 53057.22947	 &     2240    &	360.02	 \\
\hline

\end{tabular}
\end{table*}

\section{Data analysis}
\subsection{VLT Data}
The near environment of X-2 is crowded, so
in order to obtain the best possible photometric measurements, 
we used point-spread-function (PSF) fitting routines 
in {\small DAOPHOT II} (Stetson 1992), a sub-package 
of the {\small MIDAS} photometry software. 
We calibrated our absolute photometry using the exposures from 
2003 December 24 for the $B$ and $R$ bands, and 2003 December 25 for 
the $V$ band; both nights were photometric. Service mode observations
also included standard star observations of 
the Landolt fields PG0231 and SA101; our transformations from the
instrumental to the $BVR$ system are fully consistent with the zeropoints 
and color terms provided by ESO. 

\subsection{HST Data} 

Our field of interest (Fig.~\ref{hst_bvi}) 
also appears moderately crowded in the HST/ACS 
drizzled images, so we used again {\small DAOPHOT~II} for our photometric 
analysis. In order to obtain a realistic estimate of the photometric 
errors, we multiplied the pixel values by the exposure time, which gave us 
the number of detected $e^{-}$ per pixel; we then added  
the background counts subtracted by the HST pipeline. 
We selected a sample of bright, isolated stars in the field 
to model the PSF; we allowed the PSF to vary
quadratically as a function of spatial position in the frame.
We used a radius of 3 pixels for PSF fitting to the other 
stars. We carried out aperture photometry of isolated stars 
with {\small SExtractor} \citep{1996A&AS..117..393B} in order to
calculate the aperture corrections between PSF-derived brightnesses 
and those from a $0\farcs5$ aperture radius
(Table~\ref{tab_magnitudes_hst}). 
Finally we applied the values given in \citet{2005PASP..117.1049S} 
to correct between $0\farcs5$ aperture and infinite
aperture photometry. 

From the ACS instrumental magnitudes, we calculated the standard 
magnitudes in both the VEGAMAG and the Johnson-Cousins systems. 
We are aware of the limitations of such transformations, especially for
stars with peculiar spectral features. But we chose to do so in order
to facilitate a comparison between the ground-based and HST photometry. 
Although, in principle, it should not be necessary 
to transform between these sets of magnitudes \citep{2005PASP..117.1049S}, 
it turns out (as we discuss in Sect.~4.5) that the interpretation 
of the data in terms of published 
isochrones differs substantially between the (F435W, F555W) system on the one hand and the
($B, V$) system on the other. 
In order to check the transformations given by \citet{2005PASP..117.1049S} between these two
photometric systems, we used our data to derive our own calibration. 
We selected bright stars in the VLT image that are not saturated or resolved 
into multiple stars by ACS. We found that our own set of transformations 
are consistent with those of \citet{2005PASP..117.1049S}, which we use for consistency 
with other studies. We interpreted our color-magnitude diagrams in the two 
photometric systems by comparing them with evolutionary tracks and
isochrones from the Padua group \citep{1994A&AS..106..275B,2000A&AS..141..371G,2000A&A...361.1023S,2002A&A...391..195G} 
and the Geneva group \citep{2001A&A...366..538L}.

We estimated the completeness limit in all frames by adding 1000 artificial stars in each 0.2 mag bin between 24.8 mag (where the completeness is 1 in all filters) and 28.6 mag (where the completeness is 0 for all filters). For each bin we took the mean of the number of stars recovered in three runs as the completeness fraction.

\section{Results and discussion}
\subsection{Distance and metallicity of \object{NGC 1313}} 
\begin{figure*}
   \begin{tabular}{cc}
      \includegraphics[width=8.5cm, angle=90, bb=50 0 504 700]{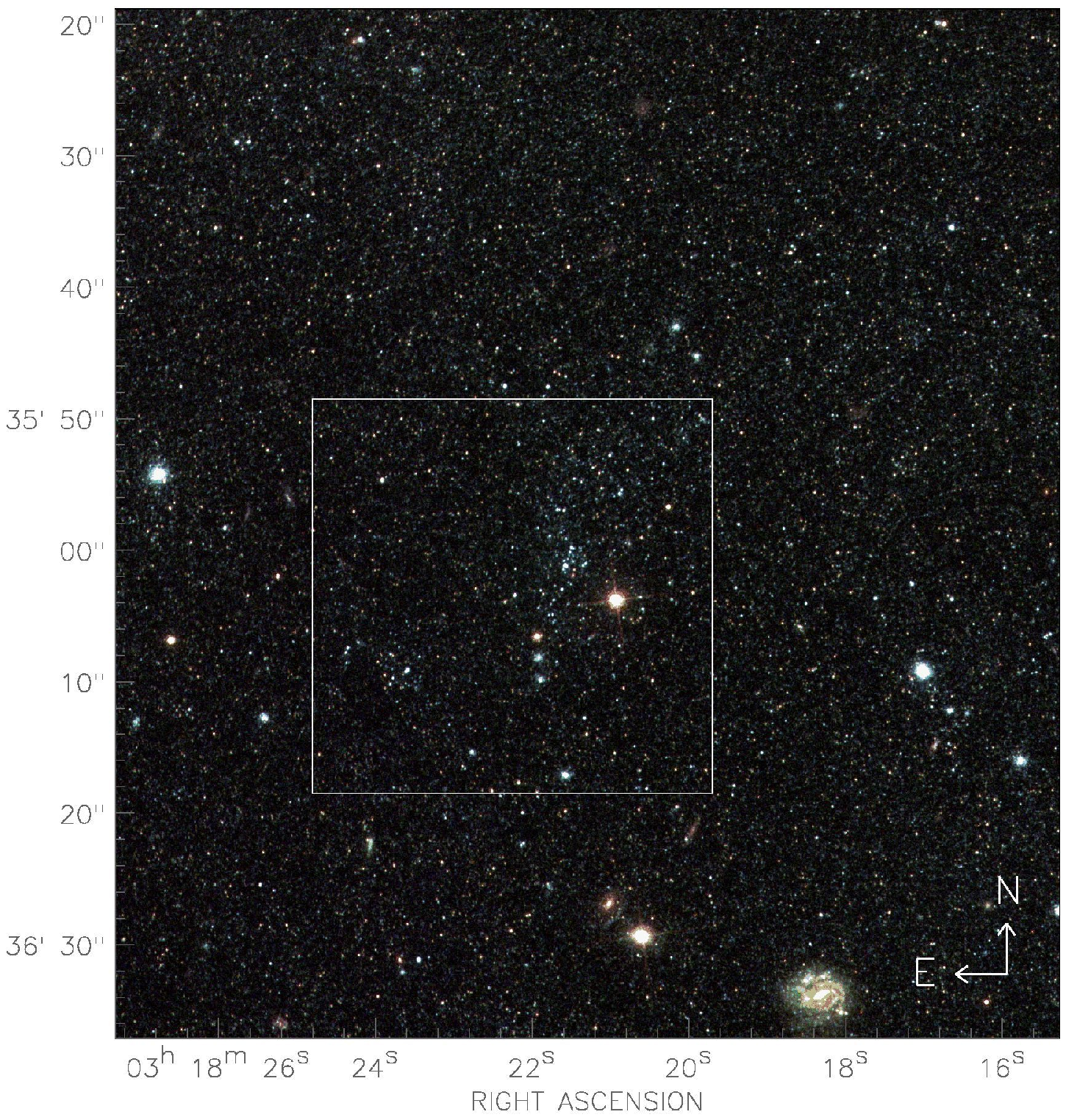}
     &\includegraphics[width=8.5cm, angle=90, bb=50 0 504 450]{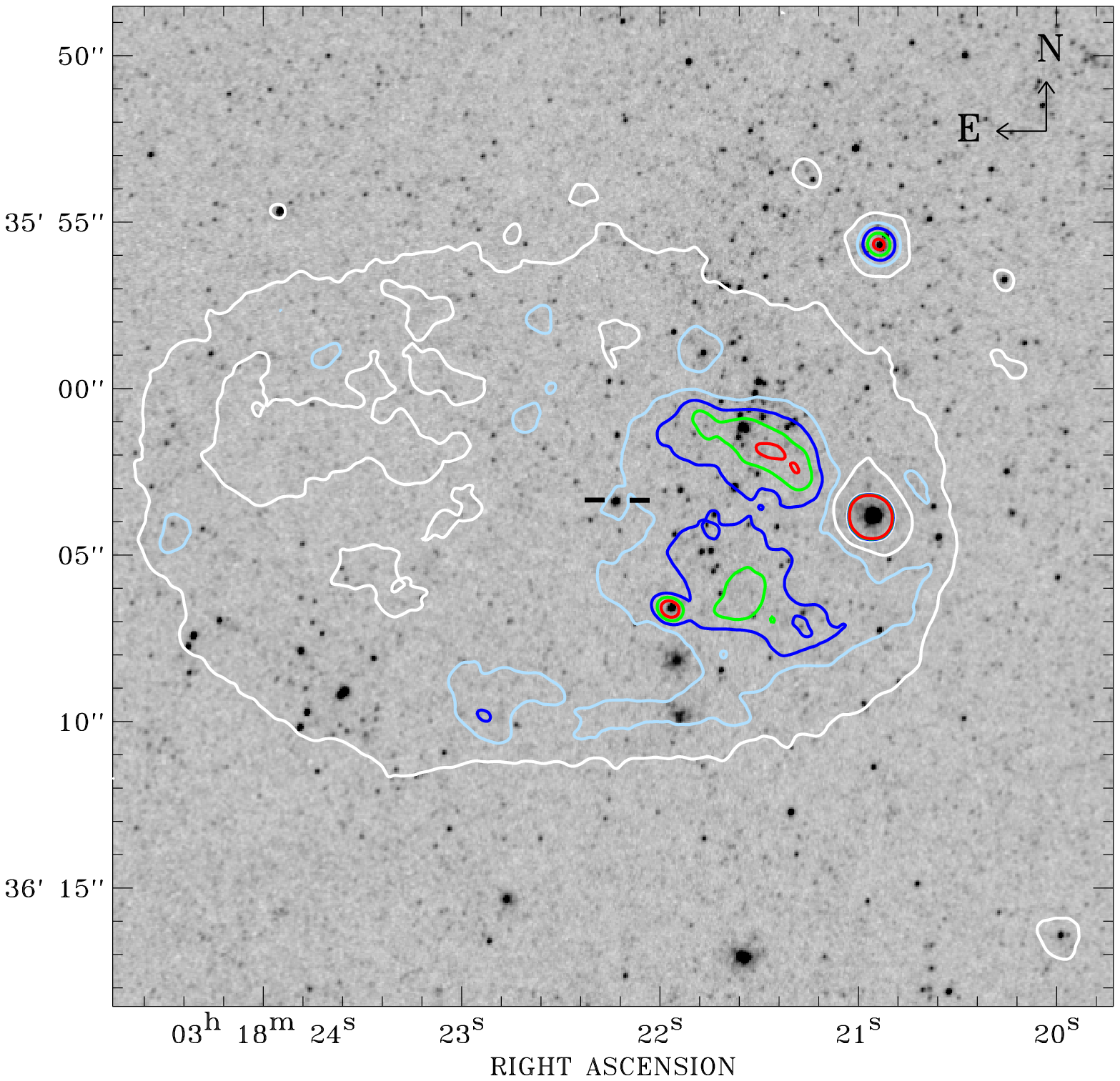}\\
   \end{tabular}
   \caption{Left panel: true color image (blue=F435W; green=F555W; red=F814W) 
of the region around X-2, from our HST/ACS observations. $1 \arcsec$ represents 19.4 pc at the distance of NGC~1313. 
The stellar environment is mainly composed of red stars from an old population. 
Two blue-star associations can be seen: the main one to the west of the ULX counterpart 
and the other (smaller) one to the south-east. Right panel: zoomed-in view 
of the immediate vicinity of the counterpart, in the F435W filter. 
The ULX counterpart is the bright point-like source at the center of the image. 
Contours of the H$_{\rm{\alpha}}$ emission (at 10, 30, 50, 70 and 90\% flux level 
above the background) are overplotted, from our VLT observations. 
Both young stellar associations are located or projected inside the H$\alpha$ 
nebula.}
   \label{hst_bvi}
\end{figure*}

\begin{figure}
      \resizebox{9cm}{!}{\includegraphics{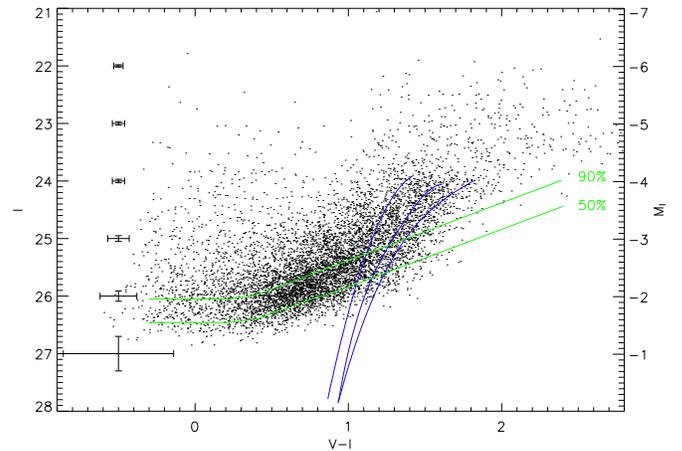}}
   \caption{$(V-I, I)$ color-magnitude diagram of all the stars located 
   in the ACS field (see Fig.~\ref{hst_bvi}, left panel). 
The estimated 50\% and 90\% completeness limits are overplotted, as are the RGB loci of three 
Galactic globular clusters of various metallicities 
(from left to right: M15, NGC6752 and NGC1851; 
data from \citealt{1990AJ....100..162D}).}
   \label{cmd_giantcurves}
\end{figure}

The HST spatial resolution is required for any stellar 
population study in NGC~1313 
(see \citealt{2002AJ....124..213M, 2007ApJ...658L..87P, 2007ApJ...661..815R} for recent work).
The central region of the galaxy is dominated by young 
stellar populations; by contrast the field around X-2 
contains few young stars and appears to be dominated 
by an older population. This is clearly shown in our ($V-I, I$) 
color-magnitude diagram (Fig.~\ref{cmd_giantcurves}) 
of the $\approx$ 7200 stars identified in the ULX neighbourhood 
(the region around X-2 shown in Fig.~\ref{hst_bvi}). 

For old stellar populations, the location of the red giant branch 
(RGB) in color-magnitude diagrams (in particular, ($V-I$, $I$)) 
is a good indicator of distance and metallicity. 
In our case, the RGB is very prominent.
We estimated the position of its tip (TRGB) 
by determining the point at which the first derivative of the $I$-band 
luminosity function has a maximum. We find that it occurs at 
$I_{0(TRGB)} \approx 24.0 \pm 0.1 \rm{mag}$ (after correcting 
for extinction). The distance can be expressed by :
$$(m-M)_0 = I_{0(TRGB)} - M_{I_{TRGB}}$$
where $(m-M)_0$ is the distance modulus and $M_{I_{TRGB}}$ is the absolute magnitude of the TRGB.
Taking $M_{I_{TRGB}}=-4.05 \pm 0.02$ \citep{2007ApJ...661..815R}, we find
that the distance modulus to NGC~1313 is $(m-M)_0 \approx 28.05 \pm 0.11$, 
corresponding to a distance of $4.07 \pm0.22$ Mpc. 
This is in agreement with the results of \citet{2002AJ....124..213M} 
and \citet{2007ApJ...661..815R} who studied the north-western region of the galaxy (distance modulus 
of $28.08 \pm 0.06$ mag and $28.15 \pm 0.03$ mag, respectively).

The intrinsic color of the RGB depends on the metal content 
of its stars \citep{1990AJ....100..162D}. A useful empirical relation
between color at $M_I = -3.5$ mag (half a magnitude below the tip) and 
metal abundance is  
$$[Fe/H] = -12.64 + 12.61 (V-I)_{0,-3.5} - 3.33 [(V-I)_{0,-3.5}]^2$$
\citep{1993ApJ...417..553L}.
To highlight the metallicity dependence, we overplotted the RGB 
loci (scaled to the distance of NGC~1313) of galactic globular clusters \citep{1990AJ....100..162D} 
with different metal abundances (Fig.~\ref{cmd_giantcurves} and Table~\ref{tab_gc}).

\begin{figure*}
   \begin{tabular}{cc}
      \resizebox{9cm}{!}{\includegraphics{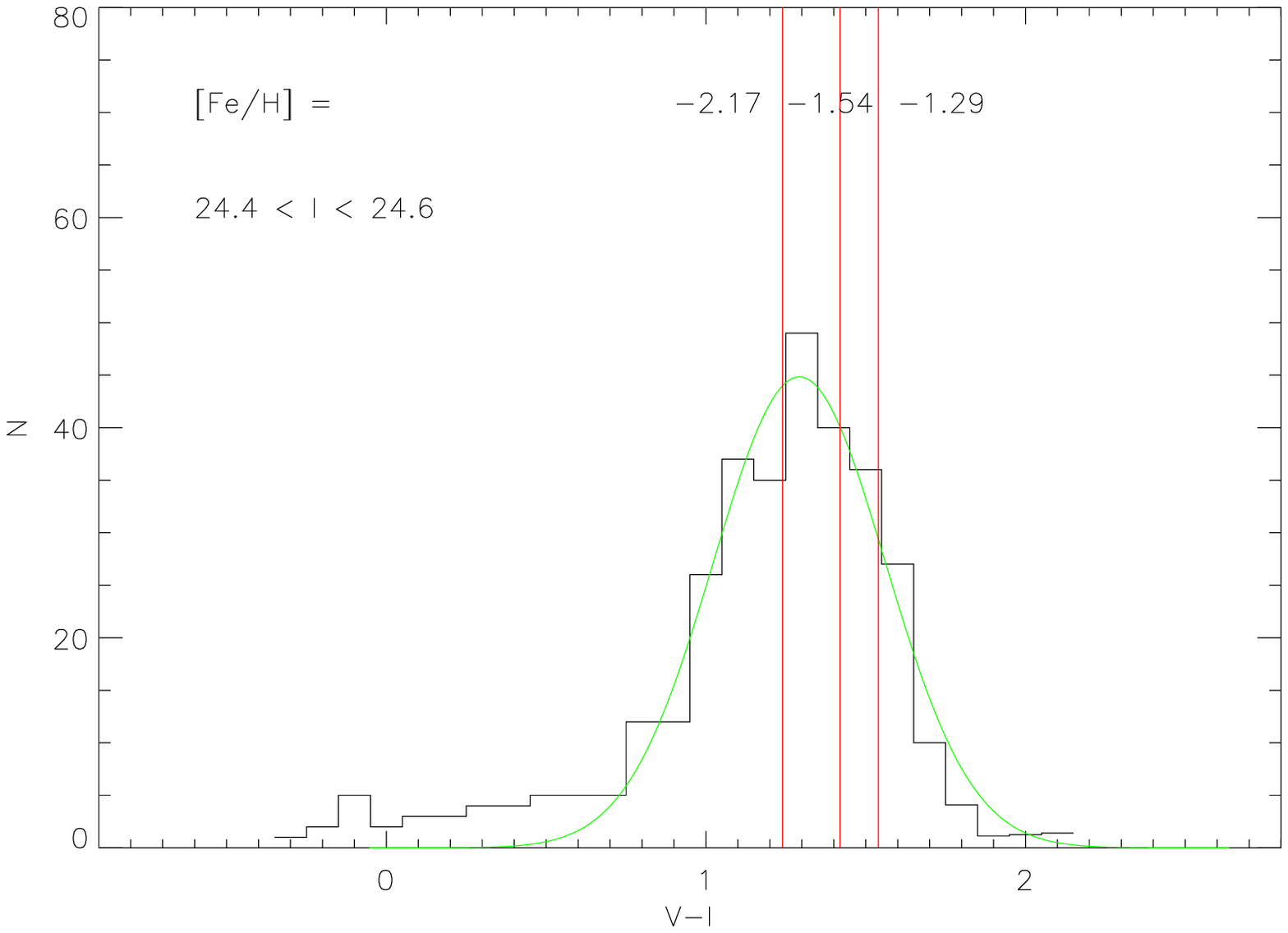}}
     &\resizebox{9cm}{!}{\includegraphics{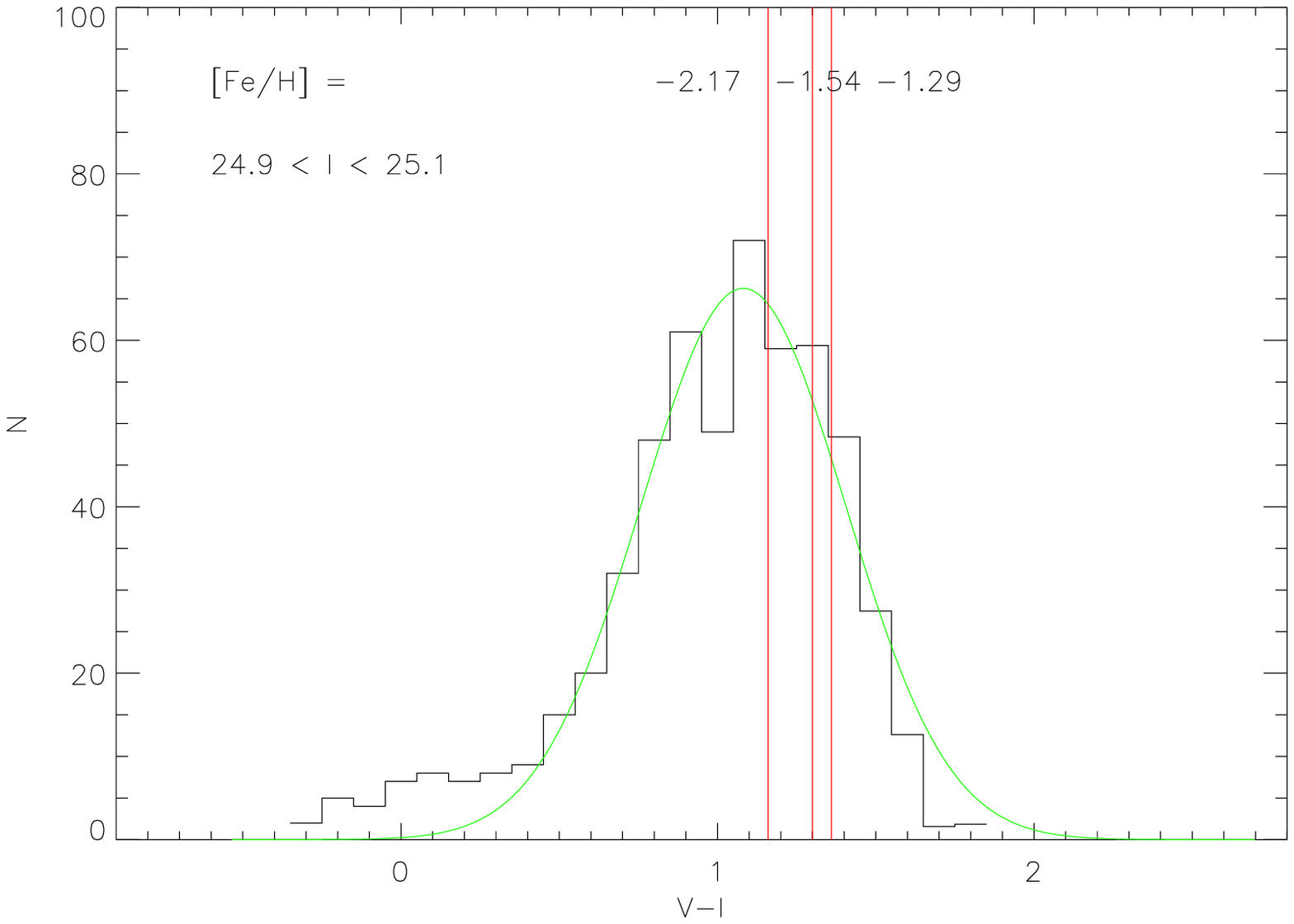}}\\
   \end{tabular}
   \caption{Color distribution of the stars in our HST/ACS field, at two values of $I$: 0.5 and 1.0 mag 
below the TRGB. The vertical lines correspond to the RGB position of three galactic globular clusters at different
   metallicities.} 
   \label{rgb_cuts}
\end{figure*}

We took horizontal cuts across the RGB at two magnitudes: 
$I = 24.5 \pm 0.1$ mag, $I = 25.0 \pm 0.1$ mag.
We determined the number distributions at those magnitudes, 
and corrected them for the completeness fraction. 
We can clearly identify the number peak in
both histograms 
(Fig.~\ref{rgb_cuts}) and their decline towards redder values of $V-I$. 
This suggests that we are seeing  the central locus of the RGB 
in NGC~1313. We estimate $(V-I)_{0,-3.5} = (1.30 \pm 0.08)$ mag 
(left panel of Fig.~\ref{rgb_cuts}). 

\begin{table}
\caption{Galactic globular clusters used for comparison with NGC~1313}
\label{tab_gc}
\centering
\begin{tabular}{lcc}
\hline\hline
Cluster	&	[Fe/H]	&	M$_{I,TRGB}$\\
\hline
NGC 7078 (M15)	& -2.17	&	-4.095 \\
NGC6752		& -1.54	&	-3.948 \\
NGC1851		& -1.29	&	-4.039 \\
\hline
\end{tabular}
\end{table}

Using the relation in \citet{1993ApJ...417..553L}, we infer 
$[Fe/H] \approx -1.9 \pm 0.3$, which is what we qualitatively expect from an old population 
at the outskirts or in the halo of a disk galaxy. 
For example, a relation between metal abundance of the halo population 
and absolute $V$ magnitude of the host galaxy was highlighted 
by \citet{2005ApJ...633..821M} (in particular, their Fig.~3).
For NGC~1313, which has $M_V = -18.72$ mag \citep{1991trcb.book.....D}, 
the expected metal abundance of its halo is $[Fe/H] \approx -2$--$-1.5$, 
in agreement with the abundance we found for the old population 
in the ULX field.

At fainter magnitudes, $M_V  \ga -2.5$ mag, the RGB is very spread out and there is a substantial 
contribution from stars with $0.5 \la (V-I)_0 \la 1.0$ (Fig.~\ref{rgb_cuts}, right panel).
They are consistent with a population of younger stars, with ages around 1--2 Gyr. 
Star formation in the field surrounding the ULX may have had various episodes 
until as recently as 1 Gyr, before the last, localized episode, responsible 
for the formation of the few young stars around the ULX 
(Fig.~\ref{hst_bvi} and Sect.~\ref{mass_age}).

\subsection{Identification of the optical counterpart}
In earlier studies of this ULX \citep{2005ApJ...633L.101M}, two possible optical counterparts, 
named C1 and C2, were suggested. A more accurate astrometric calibration 
of the {\it Chandra} and HST images pointed in favour of C1
\citep{2006ApJ...641..241R,2007ApJ...661..165L}. The two objects 
are separated by $0\farcs8$, i.e. 16 ACS pixels. More importantly and contrary to the claims of \citet{2005ApJ...633L.101M}, 
we found \citep{2006IAUS..230..293P} that they can also be clearly 
resolved in our VLT spectra with a projected distance of 4 pixels corresponding to $0\farcs5$. 
The characteristic high-excitation 
emission line spectrum displaying a broad HeII$\lambda$4686 line that we observed in C1 (\citealt{2006IAUS..230..293P}, Gris\'{e} et al, 
in preparation) provides 
the decisive proof that this is the optical counterpart of the ULX.
Since C1 is a blue stellar object (see Table~\ref{tab_magnitudes_hst}), 
this result is also in agreement with the typical 
blue colors found in most other optical ULX counterparts 
\citep{2002MNRAS.335L..67G,2002ApJ...580L..31L,2004ApJ...602..249L,2005ApJ...620L..31K,2007ApJ...661..165L}.
In fact, the optical counterpart is expected to contain emission 
from both the irradiated accretion disk and the donor star.
A possible explanation for the prevalence of blue colors 
was suggested by \citet{2007arXiv0710.3854M} and by \citet{2008MNRAS.386..543P}, based on theoretical 
evolutionary tracks of X-ray irradiated stars 
in a binary system.

\begin{table*}
\caption{Brightness of the ULX counterpart in different filters, 
from the HST/ACS observations. Magnitudes are expressed both in the HST/ACS
Vegamag system and in the Johnson-Cousins ($UBVRI$) system, when possible. When 
converting from F555W to $V$ for the 2004 Feb 22 observation, we assumed 
that the $(B-V)$ color of the counterpart was the same as on 2003 Nov 22. 
We also give the brightnesses reported by \citet{2007ApJ...661..165L}, 
that are consistent with our results except for the 
F330W/$U$ band.}
\label{tab_magnitudes_hst}
\begin{minipage}{\textwidth}
\centering
\begin{tabular}{|l|c|c|c|c|c|c|}
\hline
Filter & Exposure Time (s) & Date 	       & Aperture Correction (mag) & VEGAmag & Johnson Magnitude & VEGAmag \footnote{These
values come from \citet{2007ApJ...661..165L} and are cited for comparison with our work.}   \\
\hline
F330W / U      &     2760  & 2003 Nov. 22      & 0.885	& 21.733 $\pm$ 0.018    &   /	      	   &  22.037 $\pm$ 0.021 \\
F435W / B      &     2520  & 2003 Nov. 22      & 0.441	& 23.423 $\pm$ 0.018    &  23.49   	   &  23.470 $\pm$ 0.017   \\
F555W / V      &     1160  & 2003 Nov. 22      & 0.467	& 23.587 $\pm$ 0.032    &  23.57    	   &  23.625 $\pm$ 0.026   \\
F814W / I      &     1160  & 2003 Nov. 22      & 0.467	& 23.614 $\pm$ 0.032    &  23.61     	   &  23.640 $\pm$ 0.043   \\
F555W / V      &     2400  & 2004 Feb. 22      & 0.421	& 23.426 $\pm$ 0.037    &  $\approx$ 23.41 &  23.472 $\pm$ 0.021    \\
\hline
\end{tabular}
\end{minipage}
\end{table*}

\subsection{Groups of young stars around X-2}
\subsubsection{Masses and age}
\label{mass_age}
The multicolor image (Fig.~\ref{hst_bvi}) already suggests that the stellar 
population immediately around
X-2 is dominated by blue stars, in addition to a more uniform distribution 
of red, old stars that otherwise dominate the outskirts of the galaxy. 
The blue stars are largely concentrated in two areas separated by $15
\arcsec$ (See Fig.~\ref{hst_bvi}); the cluster located north-east 
of the ULX appears more rich in stars.

\begin{figure*}
   \begin{tabular}{cc}
      \resizebox{9cm}{!}{\includegraphics{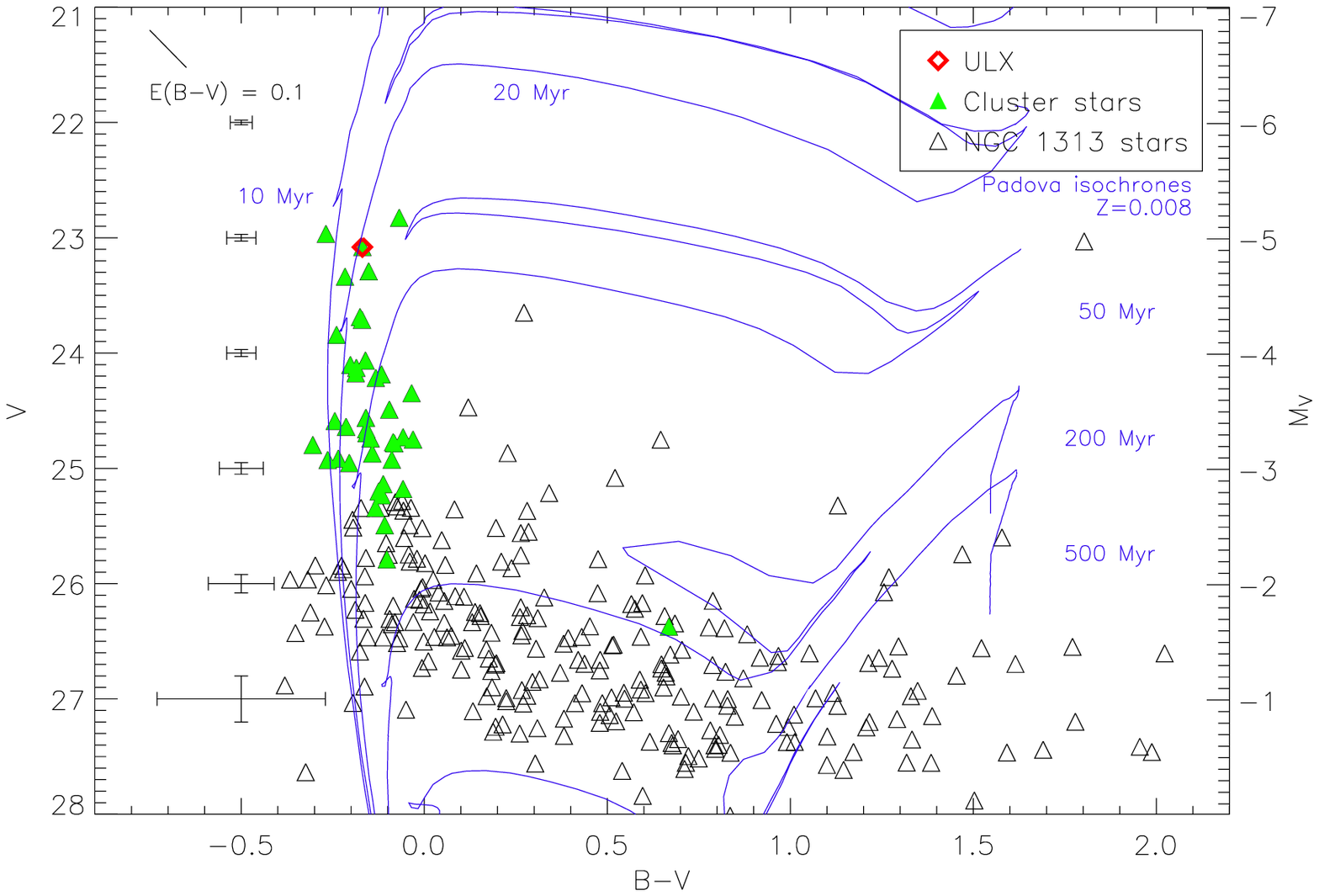}}
     &\resizebox{9cm}{!}{\includegraphics{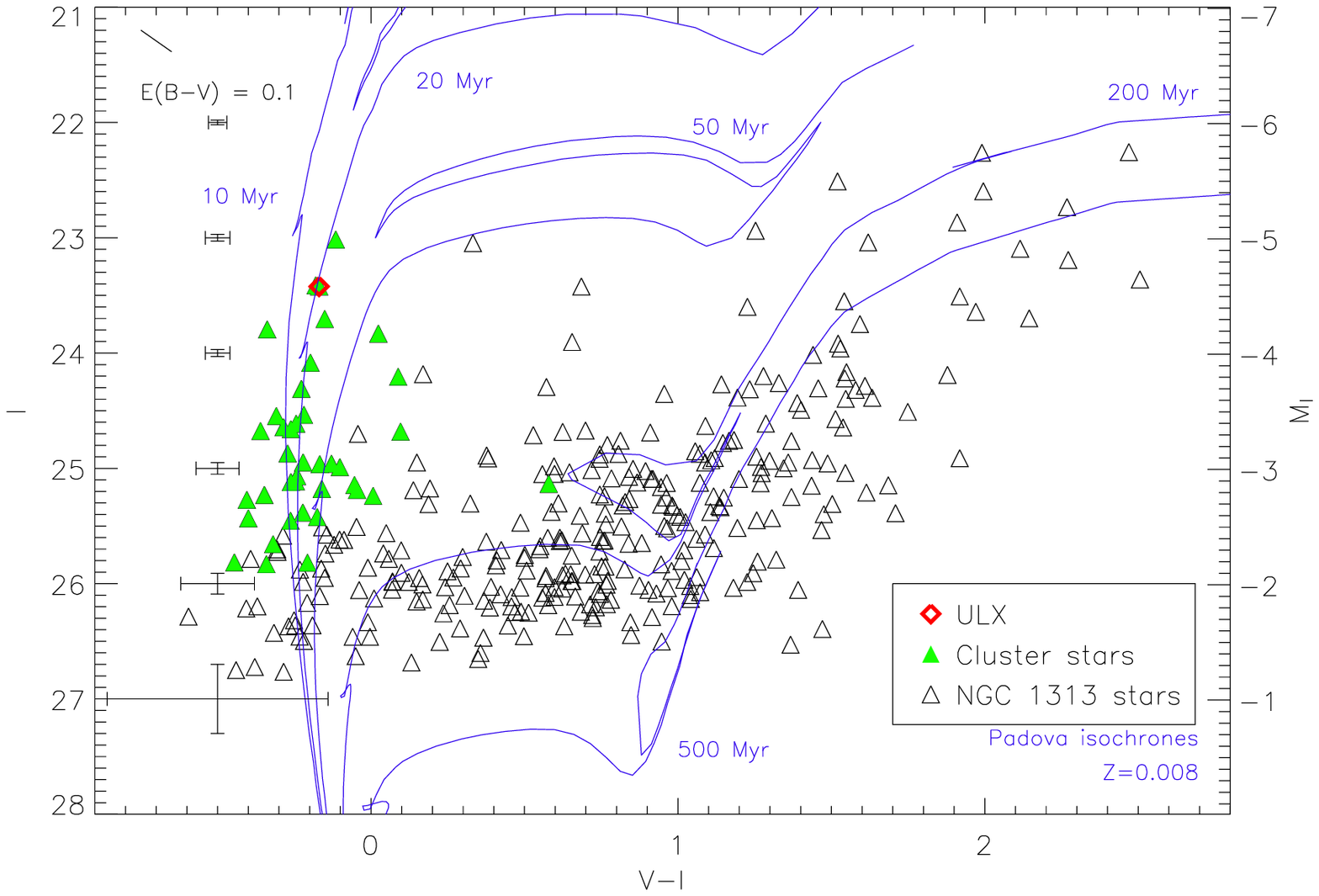}}\\
   \end{tabular}
   \caption{HST/ACS color-magnitude diagrams for the stellar field
   around the ULX. HST/ACS magnitudes were transformed into the 
   Johnson-Cousins system. Padua isochrones for stars of different ages are 
   overplotted. Typical photometric errors are also plotted. Data 
   have been corrected for the Galactic extinction (E$(B-V)
   = 0.10$ mag), the bar at the top left corner illustrating this
   effect. 
   Left panel: Color-magnitude diagram in the (B,V) system.
   Right panel: Color-magnitude diagram in the (V,I) system. The same
   isochrones are plotted in the two panels, i.e for 10, 20, 50, 200 and
   500 Myr at Z = 0.008. We can see that the two diagrams are largely
   consistent with each other, excluding the need for a high extragalactic reddening 
   $E(B-V)=0.23$ advocated by \citet{2007ApJ...661..165L}. This gives an age of 
   about 20 Myr for the brightest stars in the young cluster.} 
   \label{cmd_johnson}
\end{figure*}

\begin{figure*}
   \begin{tabular}{cc}
      \resizebox{9cm}{!}{\includegraphics{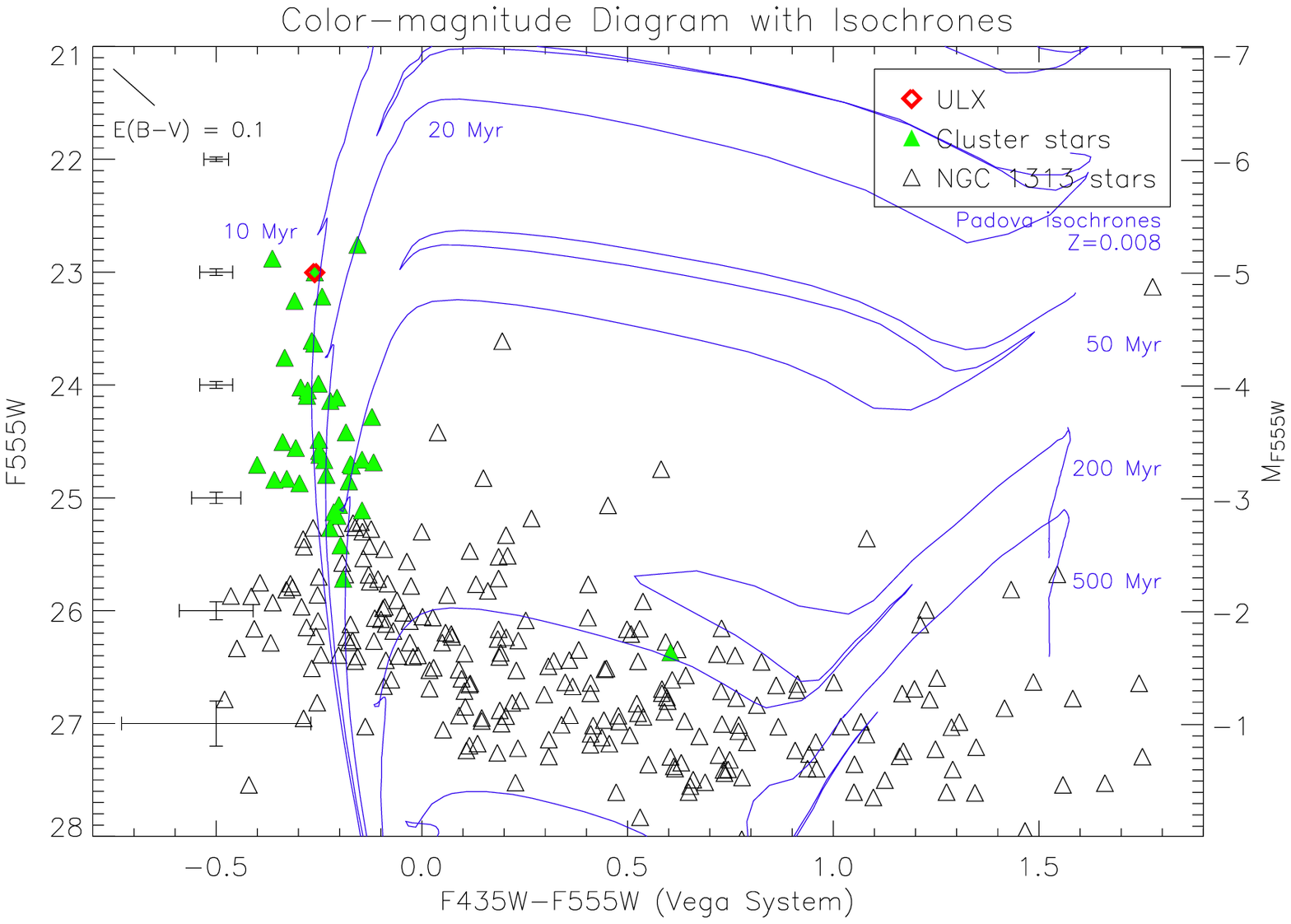}}
     &\resizebox{9cm}{!}{\includegraphics{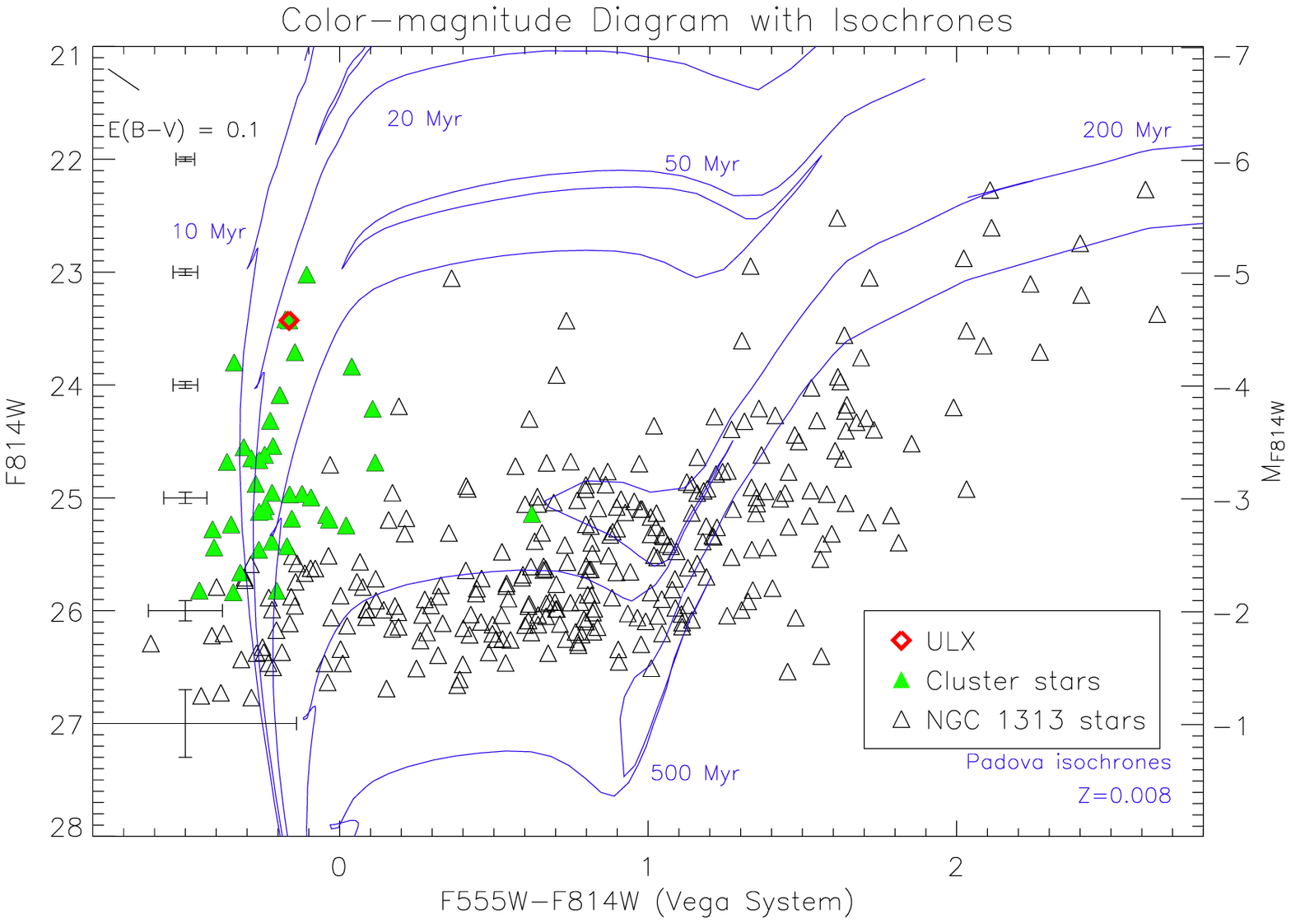}}\\
   \end{tabular}
   \caption{HST/ACS color-magnitude diagrams in the VEGAMAG
   photometry system. Left panel: color-magnitude diagram in the
   (F435W,F555W) system. Right panel: color-magnitude diagram in the
   (F555W,F814W) system. Our results are mostly consistent with those of
   \citet{2007ApJ...661..165L}; however, we argue that we need 
   a significantly lower extinction than claimed in that paper: 
   E$(B-V)=0.10$ mag instead of E$(B-V)=0.33$ mag which was used to bring the two color-magnitude diagrams into agreement.
   By comparing these diagrams with those of Fig.~\ref{cmd_johnson}, we see that the  
   ($V,I$) and (F555W,F814W) isochrones are consistent with each other. 
   However, the isochrones in the ($B,V$) and (F435W,F555W) systems are 
   shifted by $\approx 0.1$ mag in color with respect to each other. 
   We thus find that the Padua isochrones in the (F435W,F555W) system are  
   different from those in the (B,V) system, transformed with the 
   \citet{2005PASP..117.1049S} equations.}
   \label{cmd_vegamag}
\end{figure*}

\begin{figure*}
   \begin{tabular}{cc}
      \resizebox{9cm}{!}{\includegraphics{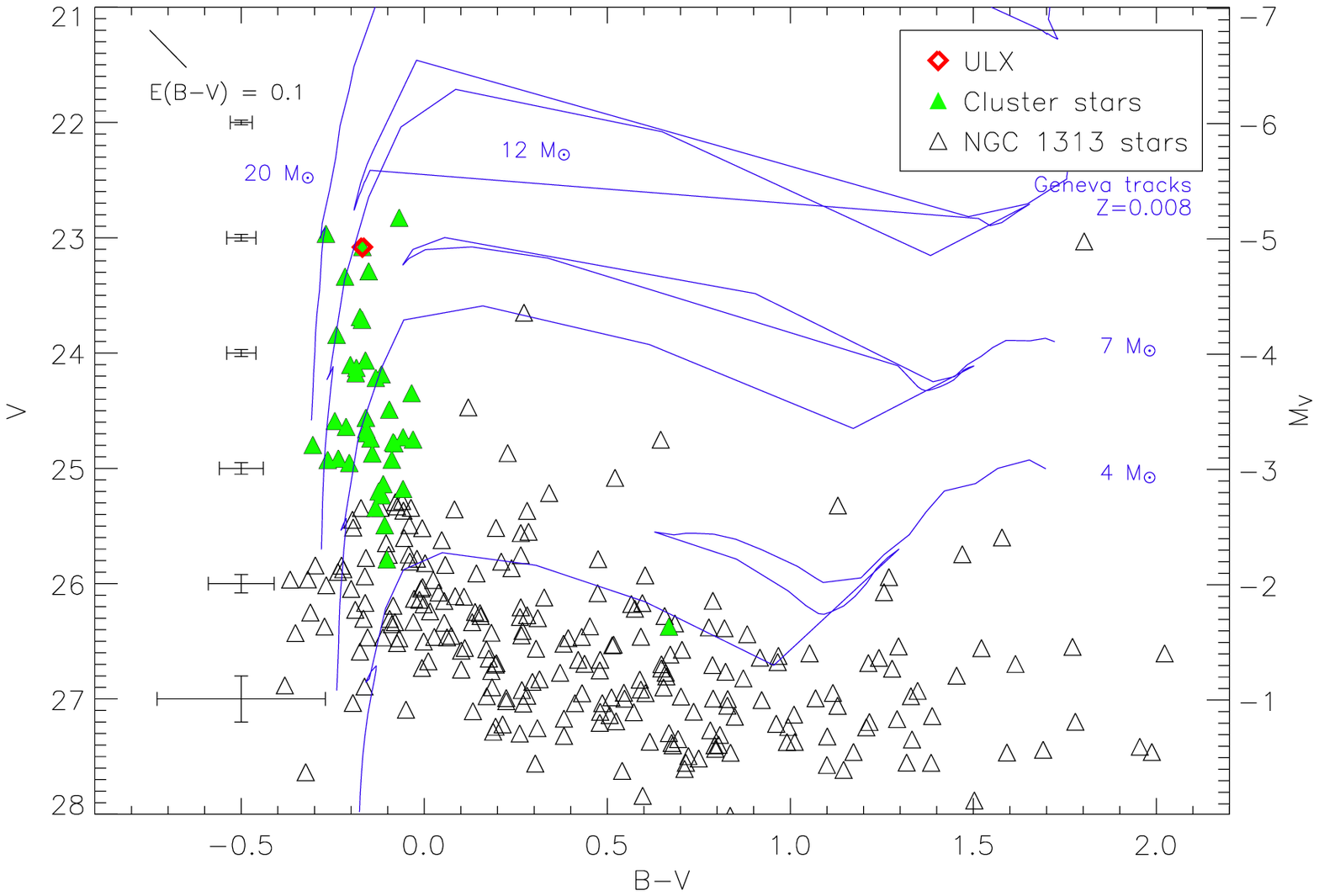}}
     &\resizebox{9cm}{!}{\includegraphics{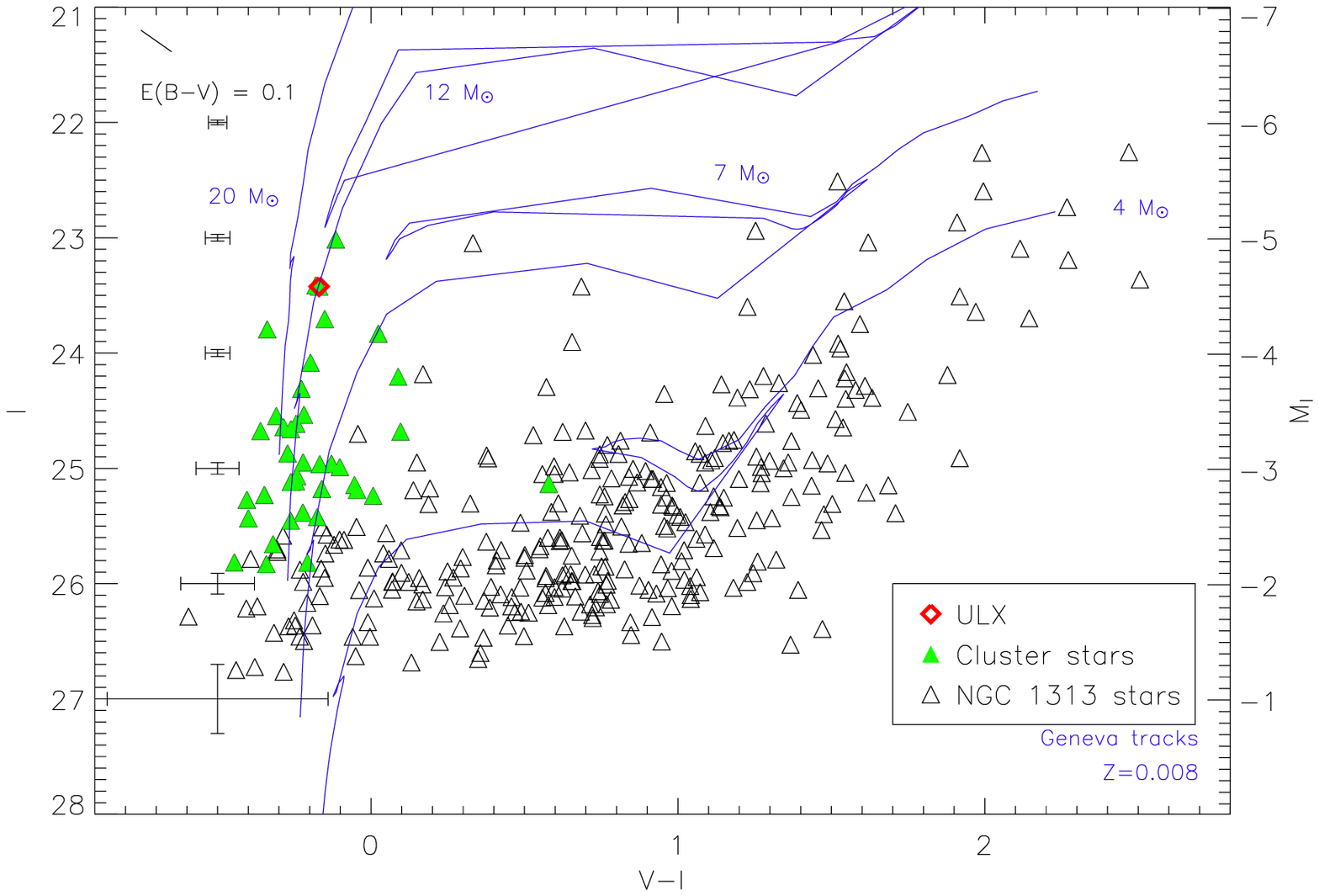}}\\
   \end{tabular}
   \caption{HST/ACS color-magnitude diagrams (Johnson-Cousins photometry)
   with Geneva evolutionary tracks for stars of different initial masses
   ($4$, $7$, $12$ and $20\ \mathrm{M_{\sun}}$) with $Z=0.008$. Left panel: color-magnitude
   diagram in the $(B,V)$ system. Right panel: color-magnitude diagram in
   the $(V,I)$ system. Both diagrams are consistent with a mass 
   $\approx 12\ \mathrm{M_{\sun}}$ for the brightest young stars in the field.} 
   \label{cmd_ET}
\end{figure*}

Color-magnitude diagrams (Figs.~\ref{cmd_johnson}, \ref{cmd_vegamag} and \ref{cmd_ET})
confirm that there are two populations of stars in the field: the
old, dominant population of the host galaxy and a blue population 
with $-0.25$~mag~$<B-V<0.0$~mag. The brightest members of the young population 
are at $V \approx 23$ mag, corresponding to $M_V \approx -5$. 
We emphasize here that all the bright stars plotted in green in the diagrams 
(Figs.~\ref{cmd_johnson}, \ref{cmd_vegamag} and \ref{cmd_ET}) were also detected in the UV band (F330W filter), which confirms 
that they belong to a young population; the red stars are not visible in the UV band.
The brightest blue stars ($V \la 24$ mag) have probably already left the main sequence, 
moving along the blue supergiant tracks.

Using isochrones from the Padua \citep{1994A&AS..106..275B,2000A&AS..141..371G,2000A&A...361.1023S,2002A&A...391..195G}  or Geneva \citep{2001A&A...366..538L} evolutionary tracks and taking into account a reddening
E(B-V)$=0.1$ and an extinction based on the Cardelli law \citep{1989ApJ...345..245C}, we derive an 
age for the cluster stars of $\approx 20 \pm 5$ Myr, which is in disagreement with the value ($\la$ 10 Myr) reported by \citet{2007ApJ...661..165L} (Sect.~\ref{photometric_system}). Our age estimate is based on the most luminous stars 
that do not suffer from large photometric errors. We used tracks corresponding to a metallicity 
$Z=0.008$, as suggested by studies of HII regions
\citep{1997MNRAS.288..726W,2007MNRAS.381..418H}; but the inferred ages are 
only weakly dependent on metallicity.
Looking at the evolutionary tracks shown in Fig.~\ref{cmd_ET}, we find 
that the track of a $12 \pm 4\ \mathrm{M_{\sun}}$ star agrees with the expected distribution of the most luminous blue stars. 
We conclude that both the mass ($12\ \mathrm{M_{\sun}}$) and brightness ($V \approx -5$ mag) 
of these stars are consistent with the inferred age of $\approx$ 20 Myr.
If on the other hand we increase the reddening to E$(B-V)=0.20$ mag 
we obtain ages around 10 Myr. However, in this case the two color-magnitude diagrams 
become less consistent and for a reddening of E$(B-V)=0.30$ mag, as proposed by \citet{2007ApJ...661..165L}, the two diagrams are no longer consistent with each other.
In fact, the color-magnitude diagram in the (B,V) system is almost non-physical at this high reddening value because a few bright stars would have a (B-V) color $\la -0.4$ mag.
Our independent estimate of the reddening (Sect.~\ref{photometric_system}) 
based on the Balmer decrement of the nebula confirms a low reddening value ($\approx 0.13$) 
towards X-2.

It may be somewhat surprising that we do not see any red supergiants, which should be 
present in a 20 Myr-old population, and are often observed in other ULX fields of 
similar age (e.g. near NGC~4559~X-1, \citealt{2005MNRAS.356...12S}).
Evolutionary tracks (e.g. Geneva tracks, \citealt{2001A&A...366..538L}) at $Z=0.008$ 
show that stars in our inferred mass range move to the red part of the diagram
for $\approx 0.4$ Myr, return onto the blue loop for $\approx 1.3$ Myr 
and then end up on the red side again for the last $\approx 0.04$ Myr 
of their lives.
We observe 5 stars brighter than $M_V = -4.5$ mag, currently evolving 
off the main sequence, towards the blue supergiant phase.
By taking a blue-to-red supergiant ratio $\sim$ 3, a value observed 
in metal-poor galaxies \citep{1995A&A...295..685L}, we expect to find about 
1 or 2 red supergiants. 
Given this small-number-statistics (coupled with the uncertainty 
in the duration of the red-supergiant phase), we do not consider 
the absence of red supergiants in the young cluster to be significant.

\subsubsection{Comparison with OB associations}
It had been known for some time that ULXs are often associated
with galaxy-wide or extended 
starburst regions (e.g. the Antennae, \citealt{2002ApJ...577..710Z}). However, some of them, 
such as Holmberg~IX~X-1 \citep{2006IAUS..230..302G,2006ApJ...641..241R} and the ULX target 
of this study, seem to be far from any {\it large-scale} star-forming activity. But the advent of 
large optical telescopes and the use of HST have now permitted the detection  
of small groups of young stars also around these ULXs, 
strengthening the association of ULXs with young stellar populations. 
X-2 is far from the center of its host galaxy, in a region where 
no recent large-scale star-formation episodes have occurred. Only the
two young groups of stars described above are present, superimposed on a predominantly
old- and intermediate-age population. These young groups are
certainly not gravitationally bound, because their density is too low
(some tens of stars scattered over 200 pc).
Based on their integrated luminosities, we infer from Starburst99 simulations \citep{1999ApJS..123....3L} that the two young 
stellar associations have masses $M \approx (5 \pm 1) \times 10^3\ \mathrm{M_{\sun}}$ 
for the north-west group and $M \approx (1.5 \pm 0.5) \times 10^3\ \mathrm{M_{\sun}}$
for the south-east group. 
So, they have similar masses and sizes as typical OB associations 
in our Galaxy and  other Local Group galaxies ($\sim 10^3\ \mathrm{M_{\sun}}$ in 
$\approx 200\ \mathrm{pc}$, 
e.g. \citealt{2003A&A...405..111G}).
The obvious question is what triggered this recent, isolated episode of star formation
in this outer part of the galaxy? It was suggested \citep[e.g. ][]{2006astro.ph.11152S} 
that  several ULXs are located in regions perturbed by tidal interactions or collisions. 
An intriguing, unexplained feature of NGC~1313 (in particular, of its southern half) 
is the presence of isolated HII regions (located $\approx 3\arcmin$ north of the ULX), 
and of some unusual, expanding HI supershells.
It was previously suggested \citep{1979AJ.....84..472S,1981MNRAS.195..451B,1994MNRAS.269.1025P} 
that the southern side of the galaxy has been affected by a collision or tidal interaction  
with a satellite galaxy. However, it was also noted \citep{1995AJ....109.1592R}
that the largest HI supershell is spatially associated with the southern HII regions.
Thus, it is also possible that the isolated HII regions and other local episodes 
of star formation are the result of collisions of large HI clouds
with the galactic disk \citep{1983A&AS...51..353M}.
It was recently noticed that such phenomena may be associated with localized star formation in NGC4395 \citep{2007arXiv0712.1184H}.
We speculate that the isolated complex of young stars around the ULX 
may also have been formed through a similar event.

\subsection{Comparison between HST and VLT photometry}
\label{comp_hst_vlt_photom}
One of the objectives of our study is to see whether there have 
been significant changes in the brightness of the ULX counterpart 
between the HST and VLT observations. To do so, we need 
to compare its brightness with those of neighboring, 
isolated stars in the two observations. This will also 
provide a check  on the absolute  photometric calibration 
between the two datasets. Such a comparison is partly hampered 
by the difference in resolution: most of the stars detected 
as single sources in the VLT images are resolved into multiple components 
 in the HST images. Besides, the brightest stars are usually saturated 
in one or the other datasets.

\begin{figure*}
   \begin{tabular}{cc}
      \resizebox{9cm}{!}{\includegraphics{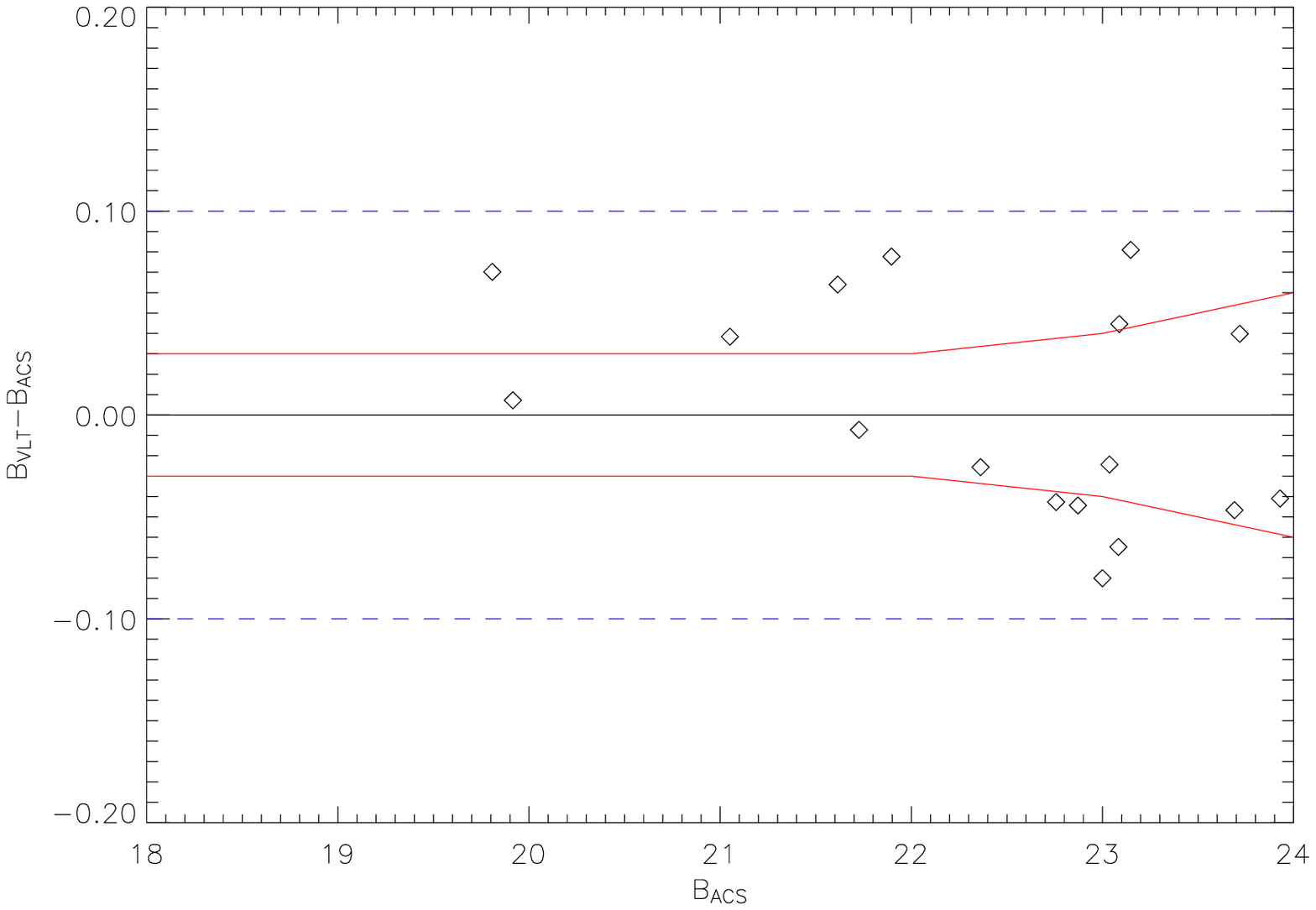}}
     &\resizebox{9cm}{!}{\includegraphics{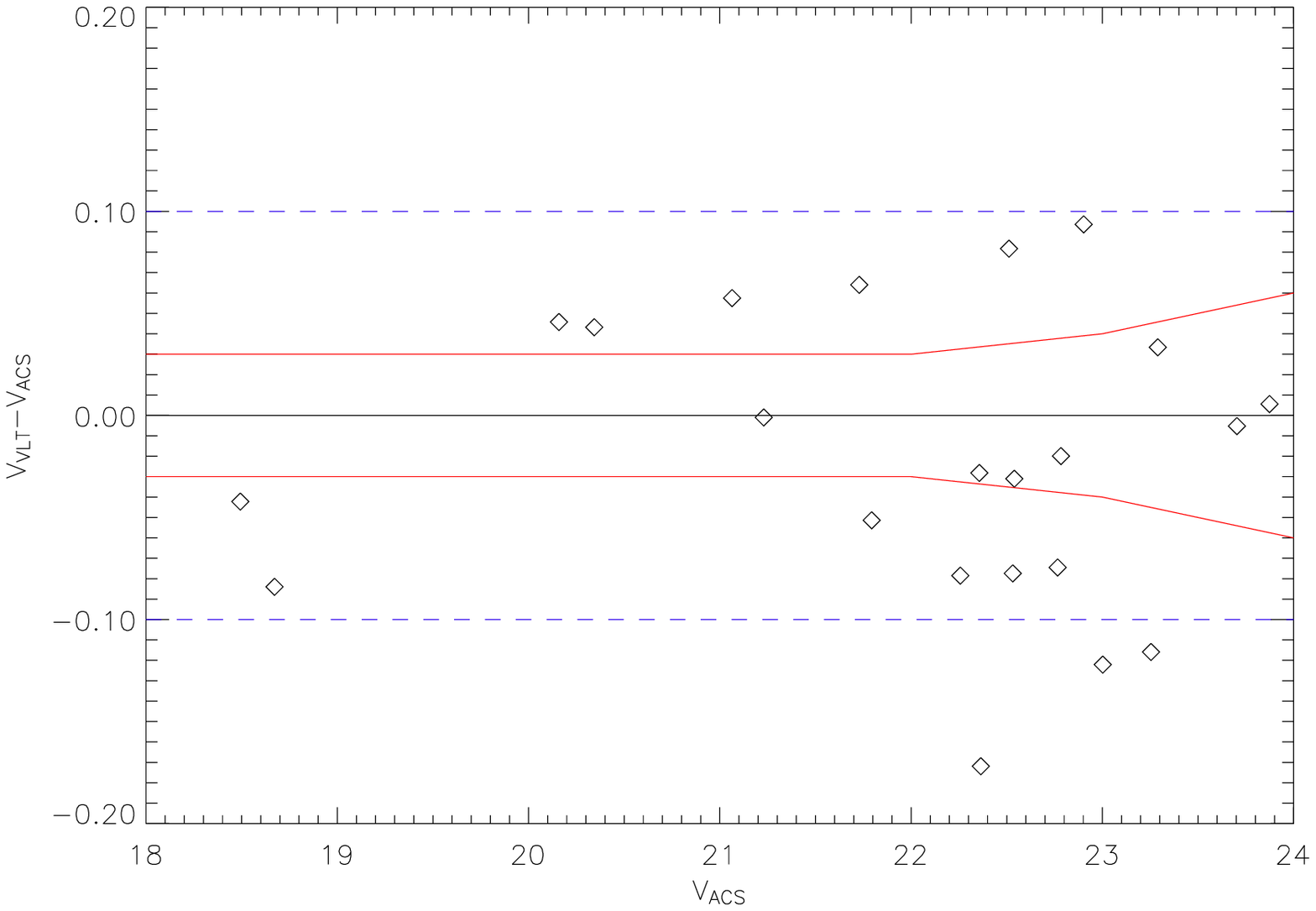}}\\
   \end{tabular}
   \caption{Difference between VLT/FORS1 and HST/ACS photometry, 
   in the $B$ (left panel) and $V$ band (right panel). For this comparison, 
   we only used bright, isolated stars. We find that 40\% of them lie 
   within the combined photometric errors of the HST and VLT observations 
   (solid red lines), and for most of the others the difference 
   is only $\le 0.1$ mag (dashed blue lines).}
   \label{comparison_hst_vlt}
\end{figure*}

Nonetheless, there are about twenty bright, isolated sources 
that appear point-like and not saturated in  either of 
the VLT and HST images. Crucially, the seeing of the VLT observations 
was good enough (Table~\ref{tab_vltdates}) to resolve the true ULX counterpart 
from its close companion.
In Fig.~\ref{comparison_hst_vlt} we show the magnitude difference of 
these stars observed with FORS1 and ACS $B$ and $V$ filters. 
For $\approx 40$\% of these stars the HST and VLT brightnesses are the same 
within the photometric errors of the two observations. But even for the other stars, 
the discrepancy does not exceed $\approx 0.1$ mag in either filter. 
Overall, the standard $1\sigma$ deviations are $\sigma_B = 0.05$ mag 
and $\sigma_V = 0.07$ mag. 
We conclude that absolute photometry between VLT and
HST can be achieved to a precision of less than about 0.1 mag.

We have then carried out a photometric study of the ULX counterpart 
in the VLT/FORS1 images (Table~\ref{tab_magnitudes_vlt}). 
Its colors could be affected by an additional error of $\approx 0.1$ mag 
due to variability in one of the filters between different observations. 
As we do not have consecutive frames in different filters, 
we used the closest observations available to minimize 
the effect of color variability. Once again, we can see that the 
absolute brightness measured in the VLT frames is in good agreement 
with the HST results. We will later show (Sect.~\ref{section_variability}) 
that the remaining differences 
could be explained as intrinsic variability of the ULX counterpart. 

\begin{table*}
\caption{Brightness of the ULX counterpart in different filters,
from our VLT/FORS1 observations.
To minimize the effects of variability, 
the $B-V$ color was estimated using the 2003 Dec 25 $B$ exposure, 
while the $B-R$ color was estimated using the 2003 Dec 24 $B$ exposure.}
\centering
\begin{tabular}{|l|c|c|c|c|}
\hline
Filter & Exposure Time (s) & Date     & Magnitude	& Absolute Magnitude\\
\hline
B      &     420/840  & 2003 Dec. 24/25   & 23.40 $\pm$ 0.02 / 23.44 $\pm$ 0.03	& -5.0\\
V      &     600      & 2003 Dec. 25      & 23.45 $\pm$ 0.04 & -4.9\\
R      &     500      & 2003 Dec. 24      & 23.58 $\pm$ 0.05 & -4.6\\
\hline

\end{tabular}
\label{tab_magnitudes_vlt}
\end{table*}

\subsection{On the conversion between the ACS photometric system and more standard colors}
\label{photometric_system}
The HST/ACS data of X-2 have also been analyzed and discussed 
in two recent papers \citep{2006ApJ...641..241R,2007ApJ...661..165L}.
We have carefully retraced their analyses and noticed some differences with respect to the 
results we have independently obtained  from the same dataset.
Most of the discrepancy seems to be related to the choice 
of photometric system for the datapoints and the evolutionary tracks employed.

We analyzed the data in two different ways which, in principle, should lead 
to identical conclusions. 
First, we compared the datapoints directly measured in the HST/ACS system colors 
with the Padua isochrones in the HST/ACS system. Then, we transformed 
the HST/ACS colors 
to the more-standard Johnson-Cousins colors via the \citet{2005PASP..117.1049S}
transformations, and compared these datapoints with the Padua Johnson-Cousins isochrones.
Both sets of isochrones were independently calculated and supplied by \citet{2002A&A...391..195G}
and Girardi\footnote{see http://pleiadi.pd.astro.it \& http://stev.oapd.inaf.it/cmd for data access} (in preparation).

Surprisingly these two methods give different results (Figs. \ref{cmd_johnson} \& \ref{cmd_vegamag}), 
leading to different estimates of stellar ages and masses~! 
To investigate the reason for this discrepancy, we also transformed the Padua Johnson-Cousins 
isochrones to the HST/ACS system using the recommended \citet{2005PASP..117.1049S}
transformations: the result is different from the Padua HST/ACS isochrones 
supplied by Girardi (in preparation), especially for the B band.

Comparing our results with those of \citet{2007ApJ...661..165L}, we can immediately see 
the effect of this discrepancy. Although our observed brightnesses and colors 
are consistent with theirs (Table~\ref{tab_magnitudes_hst}), our determination 
of the age of the young stellar population is different ($\approx 20$ Myr instead of their
$\la 10$ Myr), with obvious consequences for the physical interpretation of the ULX.
The Padua HST/ACS isochrones suggest different ages of
the young stars in the (F555W versus F435W-F555W) and 
(F814W versus F555W-F814W) color-magnitude diagrams (Fig.~\ref{cmd_vegamag}). 
An obvious way to reconcile this discrepancy is to take the younger age 
and
assume a high local extinction, which led \citet{2007ApJ...661..165L} to conclude that E$(B-V)=0.33$ mag. 

However, as we showed in Sect.~\ref{mass_age}, if we transform the observed brightnesses
and colors to the Johnson-Cousins system and compare them 
with the Padua Johnson-Cousins isochrones, we obtain the same age 
in the ($V$ versus $B-V$) and ($I$ versus $V-I$) color-magnitude diagrams (Fig.~\ref{cmd_johnson})
with no need for an extinction greater than the Galactic line-of-sight value.
An independent argument in favor of low extinction comes from the spectrum of 
the bubble nebula around the ULX, in which the Balmer decrement suggests
$E(B-V) = 0.13 \pm 0.03$ mag (Gris\'e et al., in preparation), which is 
inconsistent with the higher values advocated by \citet{2007ApJ...661..165L}.
Thus, we suspect that the Padua Johnson-Cousins isochrones are the more reliable 
set of tracks. In any case, we emphasize that the discrepancy between the two widely-used sets 
of tracks in the literature deserves further investigation. We caution that conclusions based on ACS F435W and F555W photometry alone and in combination with corresponding Padua evolutionary tracks may well be misleading.

The F435W and F555W filter data from the same HST/ACS dataset were also studied 
by \citet{2006ApJ...641..241R}. They derived an age $\ga 10^7$ yr 
and stellar masses $\la 10 \rm M_{\sun}$. Although their results appear to be in agreement 
with ours, we cannot directly compare our analysis with their approach, because 
their paper does not provide information on how they converted from 
F435W and F555W to $B$ and $V$ bands. Moreover, there seem to be some 
inconsistencies because the galactic extinction is not accounted for when 
they estimate the stellar age. Finally, \citet{2006ApJ...641..241R} found 
that the ULX counterpart has $M_V = -3.96 \pm 0.02$ mag, which 
is clearly inconsistent (by $\approx 1.0$ mag) with both our result 
and that of \citet{2007ApJ...661..165L}.

\section{Nature of the ULX optical counterpart}
\subsection{Constraints from photometry} 
We have seen (Sect.~\ref{mass_age}) that the young population of stars is consistent 
with an age of $(20 \pm 5)$~Myr, and with an upper mass limit for non-collapsed stars of
$(12 \pm 4)\ \mathrm{M_{\sun}}$. Although it is problematic to draw any inferences 
about the ULX mass-donor star from photometry alone, 
due to
unknown effects of X-ray irradiation 
and binary evolution \citep{2008MNRAS.386..543P}, 
we can at least say that the ULX optical counterpart shares 
the same brightness and colors as the brightest stars in the young association.
If the counterpart is believed to be optically dominated by a normal star, its magnitude ($M_V \sim -5$ mag) is consistent with those of an early B-type or a late-type O star, although its colors ($(B-V)_{0} \sim -0.18$, $(V-I)_{0} \sim -0.16$) are more consistent with an already somewhat evolved B-type star.
But taking into account that we also expect a contribution from the accretion disk 
to the integrated properties of the ULX counterpart (see Sect.~\ref{section_variability}), 
the estimated values of brightness and hence mass 
for the donor star are necessarily upper limits.
In fact, numerical models of stellar-mass black-hole X-ray binaries 
with donor stars in the mass range $2$--$17\ \mathrm{M_{\sun}}$ 
\citep{2005MNRAS.356..401R} suggest that the accretion disk 
should dominate the optical emission from the binary system.

From the empirical point of view, we draw attention to the well-known
\citet{1994A&A...290..133V} diagram which shows a strong
correlation over 10 absolute magnitudes between the observed quantity 
$\Sigma$ and the absolute visual magnitude M$_{\mathrm{V}}$ for disk dominated
low mass X-ray binaries. Here,  
$\Sigma = (L_{\mathrm{X}}/10^{38}\mathrm{erg/s})^{1/2}\times (P/1hr)^{2/3}\times (M/(2 \; \mathrm{M_{\sun}}))^{1/3}$
which reflects the simple assumption that the total bolometric luminosity L$_{_\mathrm{bol}}$
of an X-ray irradiated disk is proportional to the X-ray luminosity and the disk area. 
At optical wavelengths, the surface brightness varies as $T^2$,
hence $L_{\mathrm{opt}} \sim L_{\mathrm{bol}}^{1/2}\times a$, where a is the orbital separation. 
Following black hole binary evolutionary model computations by \citet{2003MNRAS.341..385P} and \citet{2005MNRAS.356..401R},
an orbital period of 1--6 days and a total mass of some $20 \; \mathrm{M_{\sun}}$ are expected for the ULX which, using 
L$_{\mathrm{X}}\sim 10^{40}$ erg/s results in $log \Sigma$ = 2.2--2.7. In the 
\citet{1994A&A...290..133V} diagram such values correspond to an absolute disk magnitude
M$_{\mathrm{V}}$ = $-4.0$ -- $-5.0$ which in fact is close to the observed brightness 
of the optical counterpart of X-2.

Taking the likely disk contribution into account, we conclude that 
the donor star in the ULX system has a mass of $\approx 10\ \mathrm{M_{\sun}}$ 
\citep{2005MNRAS.356..401R}.

With the currently available data, it is however difficult to make more 
specific statements about the companion star. More information will come from our 
spectroscopic study currently in progress\footnote{13 
hours of observations executed on VLT/FORS1},
which will constrain the orbital period and hence the size of the system, 
and will help in resolving the stellar and disk contribution 
to the optical spectrum. Deep spectra will possibly identify 
narrow absorption lines from the donor star as well 
as broader emission lines from the disk.

\subsection{Photometric variability in the VLT data}
\label{section_variability}
\begin{figure*}
   \begin{tabular}{cc}
      \includegraphics[width=9cm]{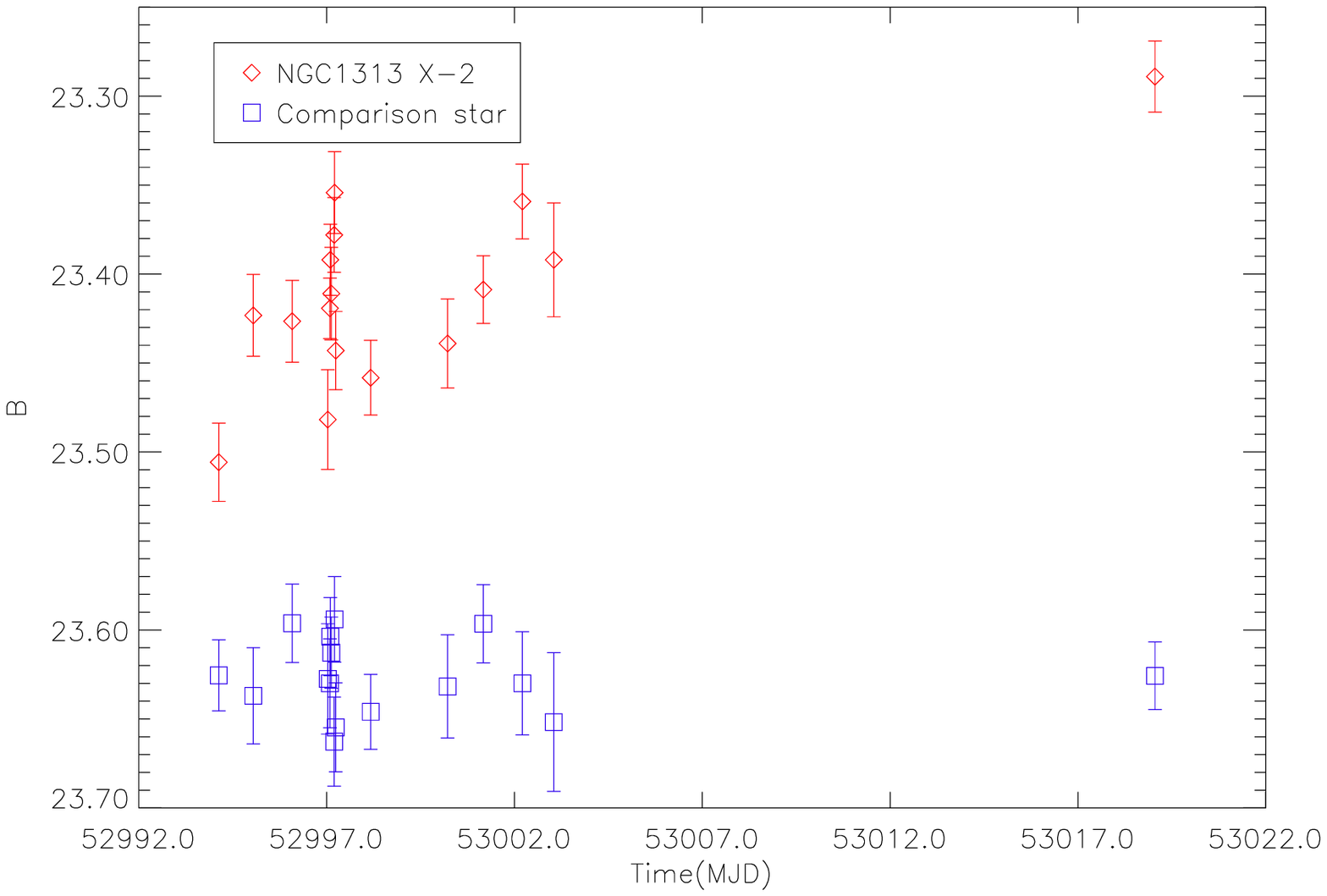}
     &\includegraphics[width=9cm]{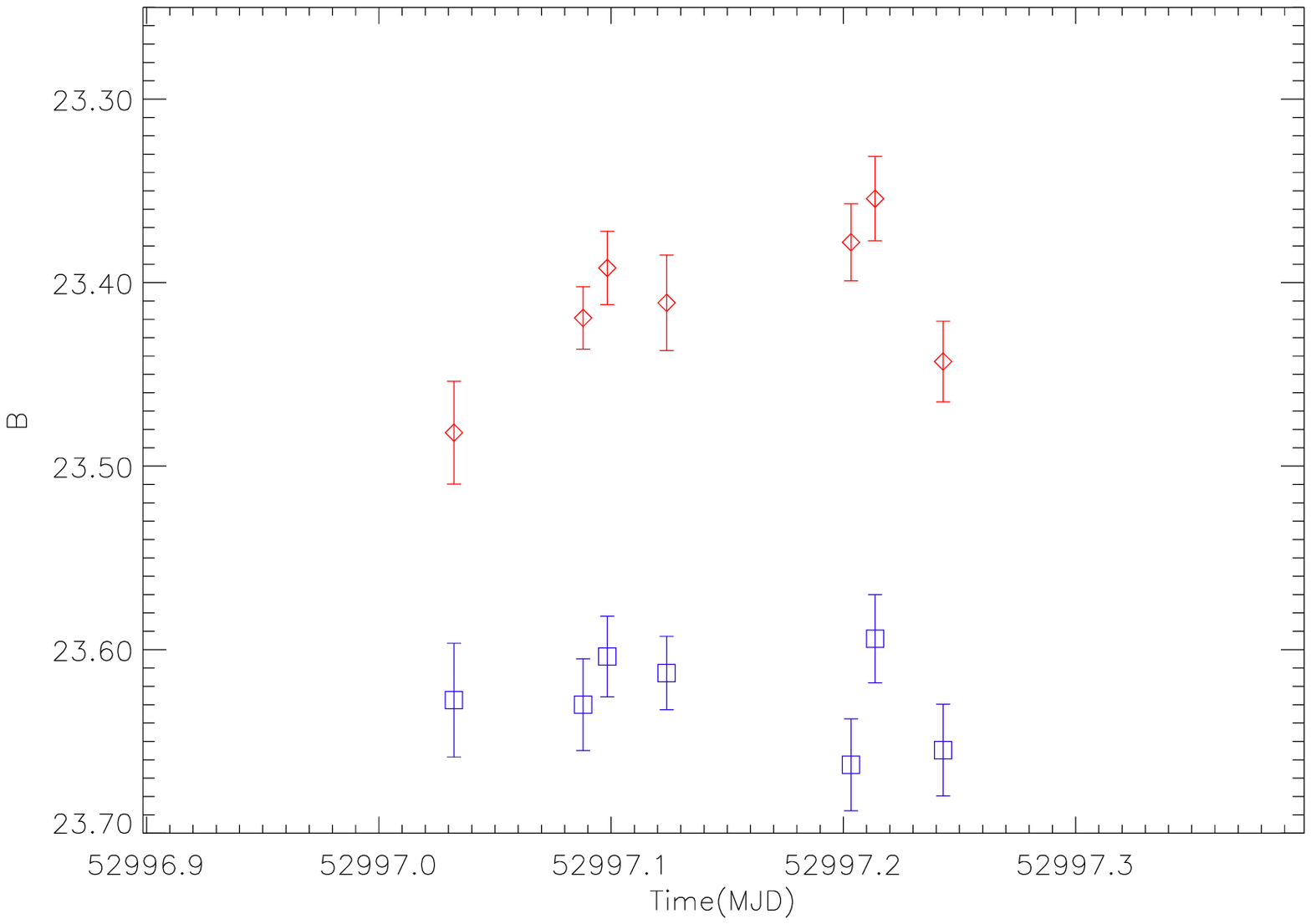}
   \end{tabular}
   \caption{Left panel: $B$-band variability of the ULX counterpart (red datapoints) 
over 1 month of observations with VLT/FORS1. The blue datapoints show the much less 
variable behavior of a comparison star of similar brightness and color; 
its zeropoint has been shifted downwards by $0.22$ mag for clarity.  
Right panel: zoomed-in view of the ULX brightness during the 2003 December 24 observations: 
the source (red datapoints) shows $\approx 0.10$ mag variability on timescales of about 
one hour. See Sect.~\ref{section_variability} for a discussion on the statistical significance of such variability.}
   \label{fig4}
\end{figure*}

In addition to determining average brightnesses and colours, 
we have searched for short and long term variability of the ULX counterpart. 
To that end we have applied differential photometry 
by using four relatively bright ($B \sim 19$--$20$ mag) comparison stars 
with relative photometric errors $<0.03$ mag. 
We have plotted (Fig.~\ref{fig4}, left panel) the $B$-band light curve 
of the counterpart and of a slightly fainter comparison (check) star of similar brightness. 
Each datapoint refers to an 840-s observation, which can be 
either a single exposure or two consecutive 420-s exposures, depending 
on the observational setup. 
The observing conditions were clear with good (sub-arcsec) seeing (Table~\ref{tab_vltdates}).
The error bars are relatively large because of the relatively large airmasses 
in many of our exposures; NGC~1313 is low in the Paranal sky 
(airmass~$\ga 1.3$). \\

\begin{figure}
   \resizebox{9cm}{!}{\includegraphics{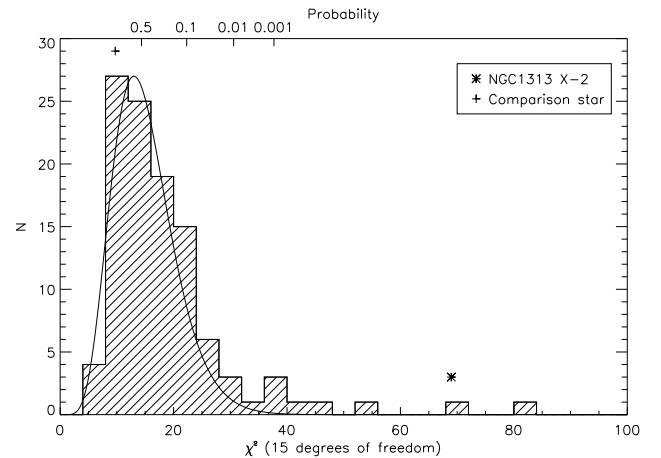}}
   \caption{$\chi^2$ test of the null hypothesis (no brightness variability) for all the stars 
in the VLT/FORS1 frame with $23$ mag $< B < 24$ mag. With a $\chi^2 \approx 84.97$ for 15 degrees of freedom (i.e, a probability of constant brightness of $7 \times 10^{-9}$), 
the ULX counterpart is more variable than almost all the other sources in the field, 
including the star we chose for comparison in Fig.~\ref{fig4}.
Overplotted is the $\chi^2$ distribution for 15 degrees of freedom.}
   \label{fig6}
\end{figure}

First, we found a trend of increasing brightness for the ULX counterpart  
(Fig.~\ref{fig4}), with a change of $\approx 0.2$ mag over 
the whole observing period. Moreover,  
repeated observations during the night of December 24 show the ULX counterpart
to also be variable on timescales of about one hour. 
By contrast, the comparison star shows significantly
less scatter, and is consistent with having a constant brightness 
within the photometric errors. 
The maximum amplitude of the variations is $0.22 \pm 0.04$ mag for 
the ULX counterpart, and only $0.06 \pm 0.05$ mag for the comparison
star. We used a $\chi^2$ test to assess more quantitatively 
whether or not the observed variability is statistically significant.
To do this, we calculated
$$
\chi^2 \equiv \sum_{i=1}^n \frac{(x_{i} - \bar{x})^2 }{\sigma_{i}^2}\mbox{,}
$$
where $n$ is the number of observations (with $n-1$ degrees of freedom),
$x_{i}$ are the measured brightnesses, $\bar{x}$ the median brightness 
and $\sigma_{i}$ the error on each measure.
Here we multiplied the formal error given by {\small DAOPHOT} by 1.15 in order to match the error distribution
of the stars in the field with the expected $\chi^2$ distribution in Fig.~\ref{fig6}.
For the ULX counterpart, we obtain $\chi^2 = 69.0$ for 15 d.o.f., 
corresponding to a 
probability of only $7 \times 10^{-9}$ for the null hypothesis 
of constant brightness. For the comparison star we obtain 
$\chi^2 = 9.7$ for 15 d.o.f, corresponding to a constant-brightness 
probability of 84\%.
During the 
night of December 24, we find $\chi^2 = 13.6$ for 6 d.o.f. for the counterpart
which translates into a probability of 3.4\% for it being constant. 
The comparison star has $\chi^2 = 5.2$ for 6 d.o.f, 
which gives a probability of 52\% for the null hypothesis.

Thus, we have 
clear
evidence that the ULX counterpart 
is variable on timescales ranging from hours to several days during our VLT observations. We further tested the significance 
of this variability by comparing the behavior of all the point-like 
sources in the brightness range $23$ mag $< B < 24$ mag. 
For 92 out of 109  sources (84\%), 
we obtain a constant-brightness probability of more than $5\%$, as expected 
from non-variable sources (Fig.~\ref{fig6}). The ULX counterpart 
clearly shows much higher scatter in its brightness measurements, 
compared with most other stars. Only 2\% of the field stars display variability equal or greater than the ULX counterpart.
In fact, a careful inspection of the position of the stars on the
frame shows that the great majority of stars with a 
probability below $0.1\%$ for the null hypothesis 
of constant brightness are located 
in the north-west sector, where crowding is more severe, 
or are situated in the immediate vicinity of other bright stars. 
In those cases, a very small fluctuation of the seeing may severely 
affect PSF fitting and brightness estimates. These stars were removed from 
the sample and are therefore not plotted in Fig.~\ref{fig6}.
Note that the ULX counterpart does not suffer from confusion 
with any other nearby sources. In conclusion, we argue 
that the observed variability of the ULX counterpart 
is statistically significant and is not due to spurious effects 
such as source confusion.

We believe that the observed scatter in the ULX brightness is
a real effect, probably due to the same kind of short-term variability 
seen in many low-mass X-ray binaries (for example, in LMC X-2: \citealt{2003MNRAS.345.1039M}). 
This is generally understood as disk reprocessing of variable X-ray emission
although simultaneous optical and X-ray observations do not always show a clear
correlation between the brightnesses in these bands.   

Other luminous stars are also known to display irregular $\sim 0.2$ mag variations in their
light curves: in particular, Be stars (e.g. \citealt{1987pbes.coll..149D}), which would also
have a luminosity consistent with this ULX counterpart. In Be stars,
the variability is mostly due to the formation of a circumstellar disk or
extended envelope, with typical radii of a few times the stellar radius
(e.g. \citealt{2003PASP..115.1153P}). However, the donor star in a ULX is thought
to be almost persistently filling its Roche lobe; accretion onto the BH
from a stellar wind or a Be disk is not sufficient to produce the observed,
persistent X-ray luminosity and would probably give rise to large X-ray outbursts.

In any case, the optical variability suggests a non-negligible flux contribution 
from the accretion disk---unlike the situation encountered in normal high-mass 
X-ray binaries, where the optical light is dominated by emission from the OB mass donor.

In order to look for periodic variations we computed a Lomb-Scargle periodogram 
of the light curve (IDL scargle routine\footnote{http://astro.uni-tuebingen.de/software/idl/aitlib/timing/scargle.html}, Joern Wilms 2000); however, no significant
period was found. 
This indicates that we are not primarily seeing ellipsoidal variations 
of the companion star over the binary orbit as would be expected by a
Roche-lobe filling star not seen pole-on. However, such variations could be masked 
by the random variations of the reprocessed X-ray flux on the accretion disk 
and on the irradiated hemisphere of the companion.

In view of this interpretation, it is useful to compare the optical and X-ray variability.
X-2 was observed by {\it XMM-Newton}  at the same time as the VLT observation.
It was found \citep{2007ApJ...658..999M} that the X-ray brightness 
rose sharply between 2003 Dec 21 and 25 (resp. MJD52294 and MJD52998) and declined 
again afterwards.
By contrast, we do not see a corresponding optical flare on a similar 
timescale (Fig.~\ref{opt_x}). The optical counterpart did not become fainter after the end 
of the X-ray flare; instead, it was $\approx 0.2$ mag 
brighter on 2004 Jan 15 than during 2003 Dec 24--25, even though 
the X-ray luminosity was lower by a factor 2.

\subsection{Variability in the combined VLT/HST data set }
\begin{figure}
   \resizebox{9cm}{!}{\includegraphics{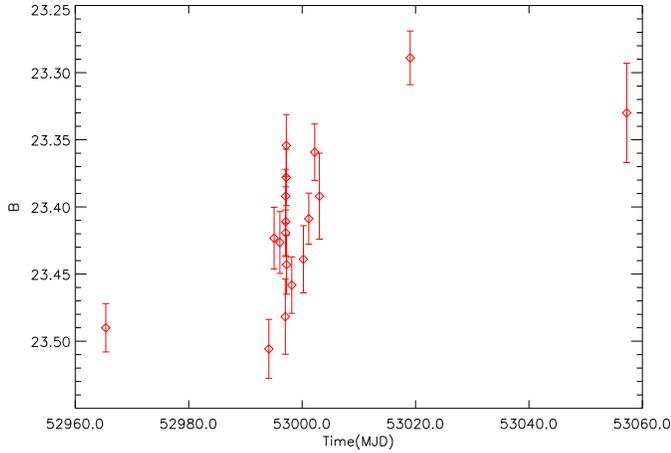}}
   \caption{Combined $B$-band light curve of the ULX counterpart between 
2003 November 22 and 2004 February 22, from the  VLT/FORS1 and HST/ACS data. 
The HST/ACS datapoints are the first and last one.}
   \label{fig5}
\end{figure}

\begin{figure}
   \resizebox{9cm}{!}{\includegraphics{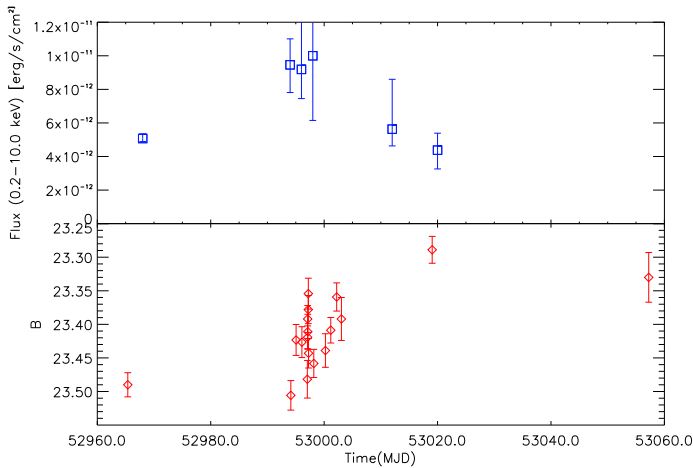}}
   \caption{X-ray (top) and $B$-band (bottom) light curve of the ULX counterpart.
Note that the X-ray flux is taken from \citet{2007ApJ...658..999M}.
}
   \label{opt_x}
\end{figure}

We showed in Sect.~\ref{comp_hst_vlt_photom} that the absolute photometry 
of our HST and VLT 
observations are consistent with each other. Therefore, we can 
extend our variability study by adding the two HST/ACS datapoints 
to the $B$-band VLT light curve discussed earlier.
We do not have an F435W HST/ACS observation on 2004 Feb 22, but 
we have used the F555W brightness and converted it to a $B$ 
brightness by using the $B-V$ color measured in the 2003 Nov HST data.
Assuming 
that the color has stayed constant, we find that the optical counterpart 
has brightened by $\approx 0.15$ mag with respect to the first HST 
observation three months earlier (Fig.~\ref{fig5}). 


\section{Conclusions}
We combined the high sensitivity of VLT/FORS1 and the high spatial resolution 
of HST/ACS for a detailed photometric study of the optical counterpart 
to the ULX X-2 and of its immediate environment in the southern part of the galaxy NGC~1313.
The dominant stellar population in that region of the galaxy is old 
($> 1$ Gyr). From the brightness and colours of its red giant branch, 
we re-estimated the distance to the galaxy as $4.07 \pm 0.22$ Mpc, 
consistent with previous studies which used observations from the north-west part of the galaxy. We also estimated that the average 
metallicity of the old stellar population is [Fe/H] $= -1.9 \pm 0.3$, 
consistent with the typical metallicity found in haloes of spiral galaxies 
of similar size.

Near the ULX, we highlighted two groups of (a few) young stars, spread out 
over $\approx 200$ pc, and hence more similar to an OB association 
(or more likely, two separate associations close to each other) 
rather than to a bound cluster. They clearly stand out in brightness and colors 
over the surrounding old population.
There are no other similar groups of young stars 
in this region of the galaxy, nor are they connected to 
spiral arm features. The reason for this recent, localized episode
of star formation is unclear, but the ULX is clearly associated 
with this young population. We speculate that the local star-formation 
episode may have been triggered by a collisional event (proposed for the Gould's Belt
in the Milky Way) with a satellite galaxy or, more likely, 
with a fast-moving HI cloud transiting across the disk plane 
(for which there is independent evidence in NGC~1313).\\

We estimate that the largest association of young stars has an age $\approx 20$ Myr 
and a stellar mass $\approx 5 \times 10^3 M_{\sun}$. 
The ULX optical counterpart appears as one of the brightest stars in the association, 
without any obvious color or brightness anomaly. Using standard stellar evolutionary 
tracks, we constrain its mass to be $\la 12 M_{\sun}$; or even less, if the accretion disk 
is significantly contributing to the source luminosity. Our mass estimate 
is smaller than reported in previous work \citep{2007ApJ...661..165L}, and correspondingly, 
we estimate relatively older ages for the young stars.
We argue that the most likely reason for that is a discrepancy 
between the Padua isochrones in the HST/ACS VEGAMAG system 
and those in the standard Johnson-Cousins system. The two sets of isochrones 
are not related to each other via the same Sirianni transformations \citep{2005PASP..117.1049S} 
that are generally accepted as the best way to transform the brightness 
and colors of the observed sources from the HST/ACS VEGAMAG system to the Johnson-Cousins system. We showed that the HST/ACS isochrones 
indicate different ages for the same dataset, in the (F814W, F555W $-$ F814W) 
and (F555, F435 $-$ F555W) color-magnitude diagrams. By contrast, 
the Johnson-Cousins isochrones give consistent ages, 
when applied to the same datapoints transformed via Sirianni's equations.
For this reason, we believe that the Johnson-Cousins isochrones 
are more reliable. In any case, this is an issue deserving
further investigation. Although the resulting differences in masses 
and ages may appear small, they can have a large effect 
on the physical interpretation of the ULX, for example regarding
its accretion rate and duration of the active phase.

One of the most significant findings of this work is the short-term 
variability of the ULX counterpart, by up to $\approx 0.2$ mag, 
on timescales of hours and days. This is detected 
both in the HST/ACS and in the VLT/FORS1 datasets, 
and even more evident in the combined dataset.
There is no evidence of periodicity. This suggests that the variability 
is not due to ellipsoidal variations. Instead, it may be caused by varying 
X-ray irradiation of the donor star and (more likely) a stochastically-varying 
contribution from the accretion disk. 

Our work, and other recent multi-band studies of ULXs, suggest that those systems share 
most of their properties with Galactic X-ray binaries, although 
at higher luminosities. The donor star need not be 
an extraordinarily massive object, or even an O star; in fact, it appears to be 
a common-or-garden B star. The accretion disk may be as luminous (or possibly even more luminous) as the 
donor star, even in the optical bands, which is consistent 
with a high X-ray luminosity and high accretion rate.
We also showed that the total stellar mass in the young stellar associations 
around the X-2 ULX is $< 10^4 M_{\odot}$: thus X-2 is clearly not in a super-star-cluster. This means that whatever the nature of the compact object is (stellar-mass or IMBH), it was not formed via a runaway-coalescence scenario as proposed by \citet{2001ApJ...562L..19E} and \citet{2002ApJ...576..899P}. But X-2 has been proposed as one of the best candidates for an IMBH ($M \sim 1000 M_{\sun}$), because this ULX has a supersoft spectral component at 0.15 keV \citep{2003ApJ...585L..37M}. We have shown in our study that X-2 sits in a normal OB association so if this ULX is an IMBH, it means that there must be other ways of forming such objects. More likely, we suggest that X-2 is not an IMBH and that it was formed within the stellar association via a more conventional stellar evolutionary scenario.
We point out here that X-2 is not an exceptional case. Indeed, most other luminous ULXs do not sit in super-star-clusters either (e.g. Holmberg~IX~X-1), but are part of much smaller associations. In those cases, either there is a different way to form IMBHs, or more likely most of them have a mass $\la 100 M_{\sun}$ and formed from ordinary stellar evolution.

For a definitive answer on the BH mass it is crucial to derive 
the parameters of the binary system, via phase-resolved observations.
The apparently random variability of the optical counterpart, interpreted as the effect 
of X-ray reprocessing in the accretion disk and on the surface 
of the secondary will probably mask periodic variability such 
as ellipsoidal variations or the X-ray heating curve. Thus, it may be hard to detect the signature 
of the binary period from photometric observations.
On the other hand, spectroscopic observations (currently in progress) 
may allow us to determine the phase-resolved radial-velocity curve 
of the HeII$\lambda 4686$ disk emission line, and hence 
to constrain the BH mass.

\bigskip

Based on observations made with ESO Telescopes at the Paranal Observatory under programme ID 072.D0614 and on observations made with the NASA/ESA Hubble Space Telescope, obtained from the data archive at the Space Telescope Institute. STScI is operated by the association of Universities for Research in Astronomy, Inc. under the NASA contract  NAS 5-26555.



\bibliographystyle{aa}
\bibliography{1313x2}



\end{document}